\documentclass[12pt]{article}

\author{\sl Derek K.\ Wise \\ \\
   Department of Mathematics\\
               University of California \\
               Riverside, CA  92521 \\
               \\
               email: derek@math.ucr.edu}
\title{Lattice $p$-Form Electromagnetism \\ and Chain Field Theory\footnote{A version with better
graphics and some related materials can be obtained at http://math.ucr.edu/$\sim$derek/pform}}
\date{July 19, 2005}

\usepackage{amssymb,amsmath}
\usepackage[all]{xy}
\usepackage{graphicx}

\newtheorem{thm}{Theorem}
\newcommand{\R}{\mathbb{R}}
\newcommand{\C}{\mathbb{C}}

\newcommand{\Rn}{\mathbb{R}^n}

\newcommand{\Z}{\mathbb{Z}}

\newcommand{\vol}{\mathrm{vol}}
\newcommand{\ran}{\mathrm{ran}\;}

\newcommand{\bfm}{\boldmath \bf}

\newcommand{\tensor}{\otimes}
\newcommand{\maps}{\colon}   
   
\newcommand{\iso}{\cong}
\newcommand{\half}{\frac{1}{2}}
\newcommand{\fourth}{\frac{1}{4}}

\newcommand{\nCob}{n{\rm Cob}}
\newcommand{\nChain}{n{\rm Chain}}
\newcommand{\Hilb}{{\rm Hilb}}

\newtheorem{defn}{Definition}

\begin{document}

\maketitle
\abstract{
Since Wilson's work on lattice gauge theory in the 1970s, discrete versions of field theories have played a vital role in fundamental physics.    But there is recent interest in certain higher dimensional analogues of gauge theory, such as $p$-form electromagnetism, including the Kalb-Ramond field in string theory, and its nonabelian generalizations.  It is desirable to discretize such `higher gauge theories' in a way analogous to lattice gauge theory, but with the fundamental geometric structures in the discretization boosted in dimension.   As a step toward studying discrete versions of more general higher gauge theories, we consider the case of $p$-form electromagnetism.  We show that discrete $p$-form electromagnetism  admits a simple algebraic description in terms of chain complexes of abelian groups.  Moreover, the model allows discrete spacetimes with quite general geometry, in contrast to the regular cubical lattices usually associated with lattice gauge theory.   After constructing a suitable model of discrete spacetime for $p$-form electromagnetism, we  quantize the theory using the Euclidean path integral formalism.  The main result is a description of $p$-form electromagnetism as a `chain field theory' --- a theory analogous to topological quantum field theory, but with chain complexes replacing manifolds.  This, in particular, gives a notion of time evolution from one `spacelike slice' of discrete spacetime to another.  
}

\newpage

\tableofcontents

\newpage 

\section{Introduction}

For the relativist, describing electromagnetism as a gauge theory has obvious appeal.  The language of fiber bundles in which gauge theory is written provides a coherent framework for doing field theory in curved spacetimes of arbitrary dimension and with very general global topology.  The bundle formalism emphasizes the idea that there is no canonical way to compare `states at $x$' with `states at $y$' when $x$ and $y$ are distinct points of spacetime --- the notion of a connection is central.  Briefly, gauge theory has much of the sort of flexibility that relativists like to demand.  

In contrast, {\em discrete} analogs of gauge theory tend to be much more rigid.  Field theories are often discretized using lattice methods \cite{Montvay}, in which Euclidean spacetime is approximated by a regular hypercubical lattice $a\Z^4$, where $a$ is some fixed length called the lattice spacing.  The lattice way of doing field theory has serious computational advantages, particularly because computer models are by nature discrete.  But lattice gauge theory  also represents a certain theoretical compromise.  As A.\ Zee has pointed out (see Chapter VII.1 of his textbook \cite{Zee}) lattice gauge theory rescues gauge invariance from the ``mangling" it suffers in perturbative field theory --- including the introduction of unphysical `ghost fields' necessary to make sense of the results --- but does so at the expense of Lorentz invariance.  The discretization shatters the symmetry of spacetime.   The lattice picks out preferred directions.  And the situation is worse for those interested in spacetimes more general than affine Minkowski space:  there is no such thing as a diffeomorphism-invariant lattice.  Of course, lattice gauge theory is often considered merely as a computational tool for doing calculations in ordinary continuum gauge theory.  One expects to recover Lorentz invariance in the limit $a\to 0$, after Wick rotation. If we are really interested in {\em manifestly} discrete theories of physics, however, the success of lattice gauge theory in its usual form is not very satisfying: we would like to have `unmangled' gauge invariance without doing such drastic damage to spacetime's symmetry.  In the context of quantum gravity, where spacetime itself is a dynamical variable, the `spin foam' approach avoids picking out preferred directions in spacetime by integrating over possible discretizations; in lattice electromagnetism we should at least like to have a theory which allows discretizations more general than the hypercubical lattices of lattice gauge theory, with the hope of recovering something that looks like diffeomorphism invariance at large scales.

More generally, we would like such a flexible discretization of {\boldmath \bf $p$-form electromagnetism}, the generalization of Maxwell's theory one obtains by replacing the gauge field $A$ by a $p$-form \cite{Carrion}.  More precisely, the gauge field in electromagnetism is a connection on a $U(1)$-bundle, which may be thought of in local coordinates as a 1-form
\[
       A = A_\mu  dx^\mu,
\]
and {\em this} may be generalized to a $p$-form:
\[
     A= A_{\mu_1\ldots \mu_p}  \frac{1}{p!} dx^{\mu_1}\wedge\cdots \wedge dx^{\mu_p}.
\] 
The electromagnetic field, locally the 2-form
\(
               F = dA
\), 
is globally just the curvature of the connection $A$.  In direct analogy, the $p$-form electromagnetic field is described locally by the $(p+1)$-form 
\[
            F= dA.
\]
The classical equations of motion for $p$-form electromagnetism, what we might call the {\boldmath \bf $p$-form Maxwell equations}, look formally identical to the 1-form case \cite{GFKG}:
\begin{align*}
       dF &= 0 \\
       \star d\star F &= J 
\end{align*} 
But here, since $F$ is a  $(p+1)$-form, $\star d \star F$ is a $p$-form, so the {\bf current} $J$ must also be a $p$-form.  One sees how the dimensions get boosted:  whereas the current in ordinary electromagnetism is fundamentally 1-dimensional, corresponding to 0-dimensional point particles tracing out 1-dimensional curves, 2-form electromagnetism has a 2-dimensional current.  One might expect such a theory to describe 1-dimensional charges sweeping out surfaces in spacetime.   It is thus not surprising that something like this shows up in string theory, where it is called the `Kalb-Ramond field' \cite{GSW,KR}.  Integrating the Kalb-Ramond field over the string's worldsheet gives a term in the action, analogous to the term in the action of a charged particle given by integrating the vector potential $A$ along the particle's worldline. 

A natural question which has been pursued by Baez and others is how to describe $p$-form electromagnetism, and other `higher gauge theories' \cite{BS, Pfeiffer} with more general gauge group, {\em globally}.  In fact, putting `$p$-form gauge theory' on the same global footing as ordinary gauge theory requires some higher-dimensional generalization of the notion of fiber bundle \cite{BS,Bartels,Bryl,MP}.  In particular, one needs to know what is meant by parallel transporting a $p$-dimensional extended object along a $(p+1)$-dimensional submanifold, rather than merely translating point particles along paths. 

Remarkably, many of the technical difficulties of defining `higher gauge theory' melt away in a discrete setting when the gauge group is abelian, that is, in the case of discrete $p$-form electromagnetism!  The need for the bundle formalism in gauge theory arises from the possibility that the gauge field might be `twisted' in a global way.  But in the lattice context, changes in the gauge field happen not continuously but in discrete steps.  It therefore becomes impossible to decide if the field is `twisted' around some noncontractible loop.  Moreover, the abelian nature of electromagnetism allows higher-dimensional generalization to be carried out in a straightforward way.  Rather than increasingly rich algebraic structures necessary for boosting the dimension in nonabelian gauge theory, higher dimensional electromagnetism can always be described in terms of chain complexes of abelian groups.  In fact, $p$-form electromagnetism on discrete spacetime turns out to be no more difficult than ordinary electromagnetism.  

It is the hope of the author that this detailed and pedagogical study of discrete $p$-form electromagnetism will serve as a starting point for further study in both lattice gauge theory and higher-dimensional analogs of gauge theories, both fields of active research, as well as theories at their interface.  A subsidiary motivation is to smooth the reader's way toward an understanding of discrete models of spacetime which arise in the study of quantum gravity, especially `spin foams'.  It is also hoped that the expository flavor of this article will make it accessible to a wider variety of readers, including physicists wishing to see an elegant treatment of lattice gauge theory and mathematicians wishing to see applications of cohomology and category theory in physics.

The plan of the paper is as follows.  {\bf Section \ref{sec:survey}} serves as an invitation to much of the material that follows it, with the intent of getting as quickly as possible to the point of doing path integral calculations in discrete quantum electromagnetism.  In this section we make an initial attempt at quantizing discrete vacuum electromagnetism with gauge group $\R$.  As typical in lattice gauge theory, we work in this section and throughout the paper in Euclidean (i.e.\ `Wick-rotated') spacetime.  Readers already well acquainted with path integrals in lattice gauge theory may wish to skim through this to fix notation and then move on to Section \ref{sec:ngraph}.  

In {\bf Section \ref{sec:ngraph}} we begin a more careful approach, considering first what mathematical model of discrete spacetime is appropriate for the description of $p$-form electromagnetism. 

{\bf Section \ref{sec:action}} draws a comparison to the continuum theory and uses this to construct an appropriate form of the action in discrete $p$-form electromagnetism.  In a subsection, it is shown that, just as ordinary gauge theories are particularly simple in two dimensions, $p$-form electromagnetism simplifies greatly in $p+1$ dimensions.  

Topological issues are considered in {\bf Section \ref{sec:cohomology}}, where the cohomology of a spacetime is introduced as a criterion for convergence of the path integral when the gauge group is $\R$.   The $p$-form analog of the Bohm-Aharonov effect is also discussed.  In the final subsection of Section \ref{sec:cohomology}, the classical theory is examined more closely, using the topological machinery developed earlier in the section.  Using Hodge's theorem, it is shown that, quite surprisingly,  under certain conditions $p$-form electromagnetism in {\em Riemannian} signature is purely topological as a classical theory.  More precisely, in the absence of charged matter the space of classical solutions depends only on the topology of spacetime. 

In {\bf Section \ref{sec:u1}} we examine the relationship between the groups $\R$ and $U(1)$ as gauge groups for electromagnetism, and use the formalism developed in previous sections for gauge group $\R$ to set up path integrals in the more usual $U(1)$ case.  These these turn out to involve not Gaussian integrals in the ordinary sense, but the analog of a Gaussian on a torus, which is nothing but a theta function.

{\bf Sections \ref{sec:zeroform}} and {\bf \ref{sec:vft}} deal with some special cases of the theory.  First, we consider $0$-form electromagnetism, which is really just scalar field theory, and find that understanding this requires a further insight in topology, namely, augmentation of complexes.  We then return to the case of $p$-form electromagnetism in $p+1$ dimensions, a case which is in fact the source of many examples in the paper.  We suggest that in $p+1$ dimensions, the theory is `almost' a topological quantum field theory --- what we call a `volumetric field theory' --- where the volume of spacetime is the only nontopological degree of freedom.  

The final section of the paper, {\bf Section \ref{sec:chain}}, presents the main result of this work, namely a description of lattice $p$-form electromagnetism that parallels the category-theoretic description of topological field theory.  The importance of this result is that it gives a rigorous description of {\em time evolution} in lattice $p$-form electromagnetism.  Namely, it allows an $n$-dimensional discrete spacetime to be split into $(n-1)$-dimensional slices representing space, and provides time evolution operators between the Hilbert spaces of states on these. 
Though some attempt has been made to retain the expository tone of the previous sections, this final section is more technically demanding than the rest of the paper.  In particular it will probably not be comprehensible without a working knowledge of categories and functors.  A highly readable introduction to categories and functors which also emphasizes applications to physics is the book by Geroch \cite{Geroch}.  The more technical aspects of category theory needed in this section are all covered in Mac Lane's textbook \cite{MacLane}.    

As this paper is intended to be expository and as self-contained as possible, an {\bf Appendix} includes a review of the Gaussian integrals needed to do calculations in discrete $p$-form electromagnetism.

\section{{\boldmath Survey: From $n$-Graphs to Path Integrals}}

\label{sec:survey}
A discrete model of spacetime might begin with a simple discrete set of points, the ``events." Soon however, if we want to write down theories that look like their desired continuum limits, we are forced to equip our model with additional data.  In the context of gauge theory, for example, we would like to be able to define an analog of the connection and take holonomies along paths, and this requires that we specify ways of getting from one event to another.  This has the effect of turning our model into a {\em graph}, with events as vertices and paths between events as finite strings of edges:
 \begin{center}
  \includegraphics
  [width=.6\textwidth]
  {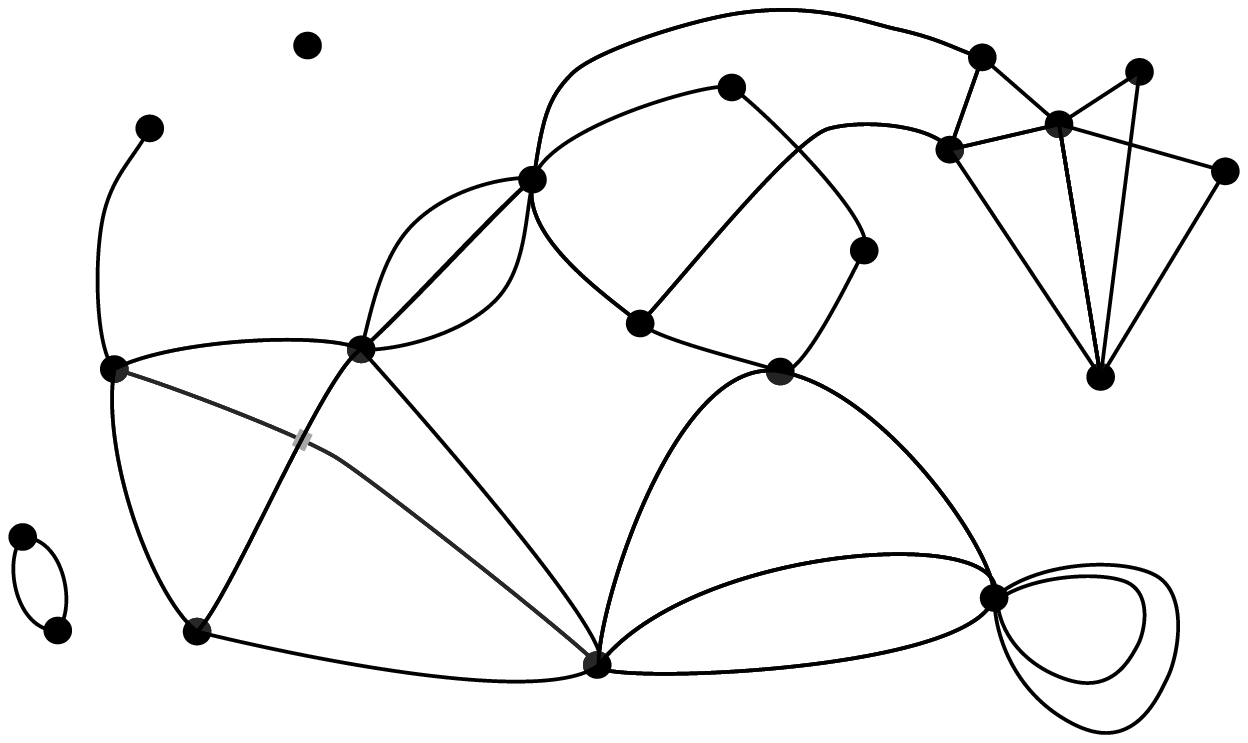}
\end{center}

Likewise, the desire to give meaning to notions such as area and curvature makes it natural to fill in empty spaces bounded by edges with 2-dimensional faces, or {\bf plaquettes}.  Thus our model becomes a `2-graph':
\begin{center}
  \includegraphics
  [width=.6\textwidth]
  {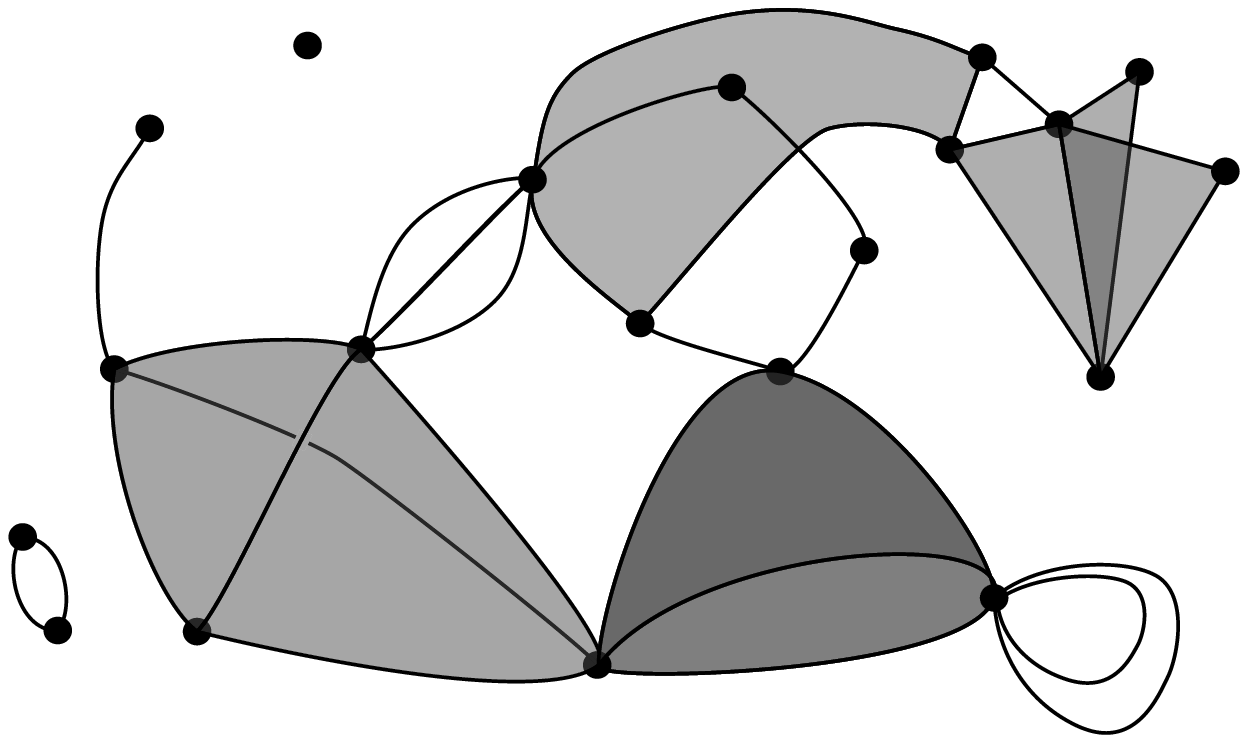}
\end{center}

In a similar way, we are led to adjoin 3-dimensional cells bounded by plaquettes, and so on, all the way up to the highest dimension of the spacetime.  In general, an $n$-dimensional discrete spacetime should be constructed from $n+1$ distinct classes of geometric objects: vertices, edges between vertices, plaquettes bounded by edges, 3-cells bounded by plaquettes, ..., and $n$-cells bounded by $(n-1)$-cells.  Spacetime thus becomes an `$n$-graph'.

The problem is, nobody knows what an $n$-graph is!  Of course, we have some solid ideas of what an $n$-graph should be like, but there seems to be no accepted intrinsic definition that is flexible enough to encompass every would-be example.  We take up this issue in Section~\ref{sec:ngraph}. 

For now, let us rely on our intuitive notion of a 2-graph as a combinatorial object consisting of finite sets of vertices, edges, and plaquettes, and begin with a na\"ive attempt at discretizing quantum electromagnetism in 2 dimensions.  Let $V$ be the set of {\bf vertices}, $E$ the set of {\bf edges}, and $P$ the set of {\bf plaquettes} or {\bf faces}.  An edge is just a line segment or curve connecting one vertex to another, while a plaquette is a 2-dimensional surface whose boundary consists of one or more edges.  A simple example, with 2 vertices, 3 edges, and 2 plaquettes, is depicted below.
 \[
\xy
(0,0)*{\includegraphics{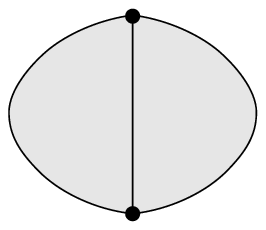}};
\endxy
\]

In practice, it is helpful to arbitrarily assign an orientation to each cell.  It is also useful to give labels to cells, just as it is sometimes convenient to use coordinates in ordinary gauge theory.  We shall label the elements of $V$, $E$, and $P$ using lower case letters $v$, $e$, and $p$, with subscripts to distinguish between them, as shown below:
\newsavebox{\twoface}
\savebox{\twoface}{
\xy
(0,0)*{\includegraphics{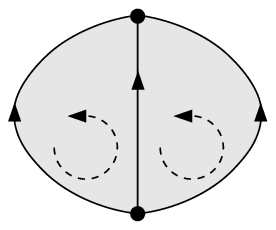}};
(0,-10)*{}="v1";
"v1"+(0,-3)*{\displaystyle v_1};
(0,10)*{}="v2";
"v2"+(0,3)*{\displaystyle v_2};
(-15,0)*{e_1};(-2.5,3)*{e_2};(15,0)*{e_3};
(-5,-3)*{p_1};(5,-3)*{p_2};
\endxy}
\[
\usebox{\twoface}
\]

The usual gauge group for electromagnetism is $U(1)$, but there is another obvious choice, namely $\R$.  We will follow the latter alternative in the first part of the paper, both because a noncompact gauge group leads to some instructive problems, and because the construction of the $U(1)$ path-integral in Section \ref{sec:u1} is made easier having established the formalism with gauge group $\R$.  In lattice gauge theory, the {\bf gauge field} or {\bf connection} $A$ assigns to each edge in the lattice an element of the gauge group:
\[
       A\maps E\rightarrow\R.
\]
The space of connections is therefore just 
\[
{\cal A} = \R^E.  
\]

Now suppose we move a test particle from one vertex to another along some path, i.e.\ along some concatenation $\gamma=e_1^{\varepsilon_1} e_2^{\varepsilon_2} \ldots e_n^{\varepsilon_n}$ of oriented edges,  where $\varepsilon_i=\pm 1$ with $e_i^{-1}$ denoting the edge $e_i=e_i^{+1}$ with its orientation reversed.  We define the {\bf holonomy} of $A$ along the path $\gamma$ to be the sum
\[
H(A,\gamma)=\varepsilon_1 A(e_1) + \varepsilon_2 A(e_2) + \cdots + 
\varepsilon_n A(e_n)
\]
Note that we use addition and subtraction because $\R$ is an additive group.  Most gauge theories, including the $U(1)$-version of electromagnetism, use multiplicative groups, so the holonomy becomes $H(A,\gamma)= A(e_1)^{\varepsilon_1}A(e_2)^{\varepsilon_2}\cdots  
A(e_n)^{\varepsilon_n}$ instead.

In gauge theory, one way to tell if a connection is curved is to send a test particle around some small loop and keep track of what happens to the connection as it goes around.  In the context of $U(1)$ electromagnetism, this means comparing the phase of the particle before and after its trip around the loop.  In the present context, our gauge group is not the group of phases, but the idea is similar.  Accordingly, we define the {\bf curvature} or {\bf field strength} $F(p)$ on a plaquette $p\in P$ to be the holonomy of the connection $A$ around the loop that forms its boundary $\partial p$.  
For example, in the lattice
\[
\usebox{\twoface}
\]
 the curvature on plaquette $p_1$ is $A_2-A_1$, while the curvature on $p_2$ is $A_3-A_2$, where  $A_i:=A(e_i)$.  Note that since the gauge group is abelian, the holonomy around the loop does not depend on what order we traverse the edges in!  This fact will be of great use to us.

To quantize our theory, we will use the Euclidean path integral approach, and to do this we must specify the form of the action for the gauge field.  Roughly, when there are no sources present, the action should be a positive quadratic function of the curvature, so that the connections with zero curvature have the least action, hence are most probable.  In Section \ref{sec:action}  we discuss in more detail what the action for a given connection should look like.  In the case of lattice electromagnetism in two dimensions, we find that the best choice for the action is  
\[
S(A)=\frac{1}{2e^2}\sum_{p_i \in P} \frac{{F_i}^2}{V_i},
\]
where $F_i=F(p_i)$ is the curvature induced by the connection on the $i$th plaquette, $V_i$ is the area of the $i$th plaquette, and $e^2$ is the square of the charge on the electron (or if you wish, the fine structure constant $\alpha=e^2/\hbar c$, since we are using units with $\hbar=c=1$). 

An {\bf observable}  $O$ in our theory is a real-valued function of the gauge field:
\[
O\maps {\cal A}\rightarrow \R
\]
so we can try to calculate its expected value, using our action $S(A)$, as
\[
\langle O \rangle = \frac{\displaystyle\int_{\cal A} O(A)e^{-S(A)} {\cal D}A}
                         {\displaystyle\int_{\cal A} e^{-S(A)} {\cal D}A}. 
\]
In certain cases, it might be difficult to define exactly what this means.  For example, if our lattice is infinite, then we must deal with the difficulties of defining measures on infinite dimensional Euclidean space.  However, even in the finite dimensional case, where ${\cal D}A=d^E\!\! A$ is just Lebesgue measure on ${\cal A}=\R^E$, when we try to calculate $\langle O \rangle$ by the above formula we usually run into serious problems!  To see this, let us do a sample calculation of the integral in the denominator, which is just the \textbf{partition function}
\[
Z:=\int_{\cal A} e^{-S(A)} {\cal D}A
\]
against which all other expected values are normalized.

For convenience, let us take the little two-plaquette complex above as our spacetime, with plaquette areas $V_1$ and $V_2$.  We then write the action as a function of the connection using matrix notation:
\begin{align*}
S(A) &= \frac{1}{2e^2}\left(\frac{{F_1}^2}{V_1} + \frac{{F_2}^2}{V_2}\right)\\
     &= \frac{1}{2e^2}\left(\frac{(A_1 - A_2)^2}{V_1}+\frac{(A_2- A_3)^2}{V_2}\right) \\
     &=\frac{1}{2e^2V_1V_2}
\begin{bmatrix}
  A_1 & A_2 & A_3 
\end{bmatrix}
\begin{bmatrix}
  V_2 & -V_2 &  0 \\
 -V_2 &  V_1+V_2 & -V_1 \\
  0 & -V_1 &  V_1 
\end{bmatrix}
\begin{bmatrix}
  A_1 \\ A_2 \\ A_3 
\end{bmatrix}\\
&=\frac{1}{2e^2V_1V_2}\langle A, QA \rangle
\end{align*}
where $A$ in the last line denotes the vector with components $(A_1, A_2, A_3)$ and $Q$ is the $3\times 3$ matrix in the line above. 
If we na\"ively apply one of the integal formulas derived in the Appendix, we obtain
\begin{align*}
  Z= \int_{\R^E} 
         e^{-\frac{1}{2e^2} \langle A, Q A \rangle} d^{E}\!\!A
   &=\sqrt{\frac{(2\pi e^2 V_1 V_2)^3}{\det(Q)}}.
\end{align*}
However, when we actually calculate the determinant we find $\det (Q)=0$, thereby obtaining the most famous wrong answer in quantum field theory: $Z=\infty$!  This is obviously bad.  

Let us examine {\em why} the integral diverges.  Notice that if we transform the connection as follows:
\[
   (A_1,A_2,A_3)\mapsto (A_1+d\phi,A_2+d\phi,A_3+d\phi),
\]
where $d\phi$ is any real number, the action
\[
    S(A)=\frac{1}{2e^2}\left(\frac{(A_1 - A_2)^2}{V_1}+\frac{(A_2- A_3)^2}{V_2}\right)
\]
is unaffected. This symmetry of the action is a degree of `gauge freedom'.  From this perspective, the divergence of the partition function should hardly be surprising --- the integrand $\exp(-S)$ is equal to $1$ along an entire line through the origin of $\R^E$!    More pictorially, one cross section of $\exp(-S)$ looks like this:
\begin{center}
  \includegraphics[width=2 in,height=1in]{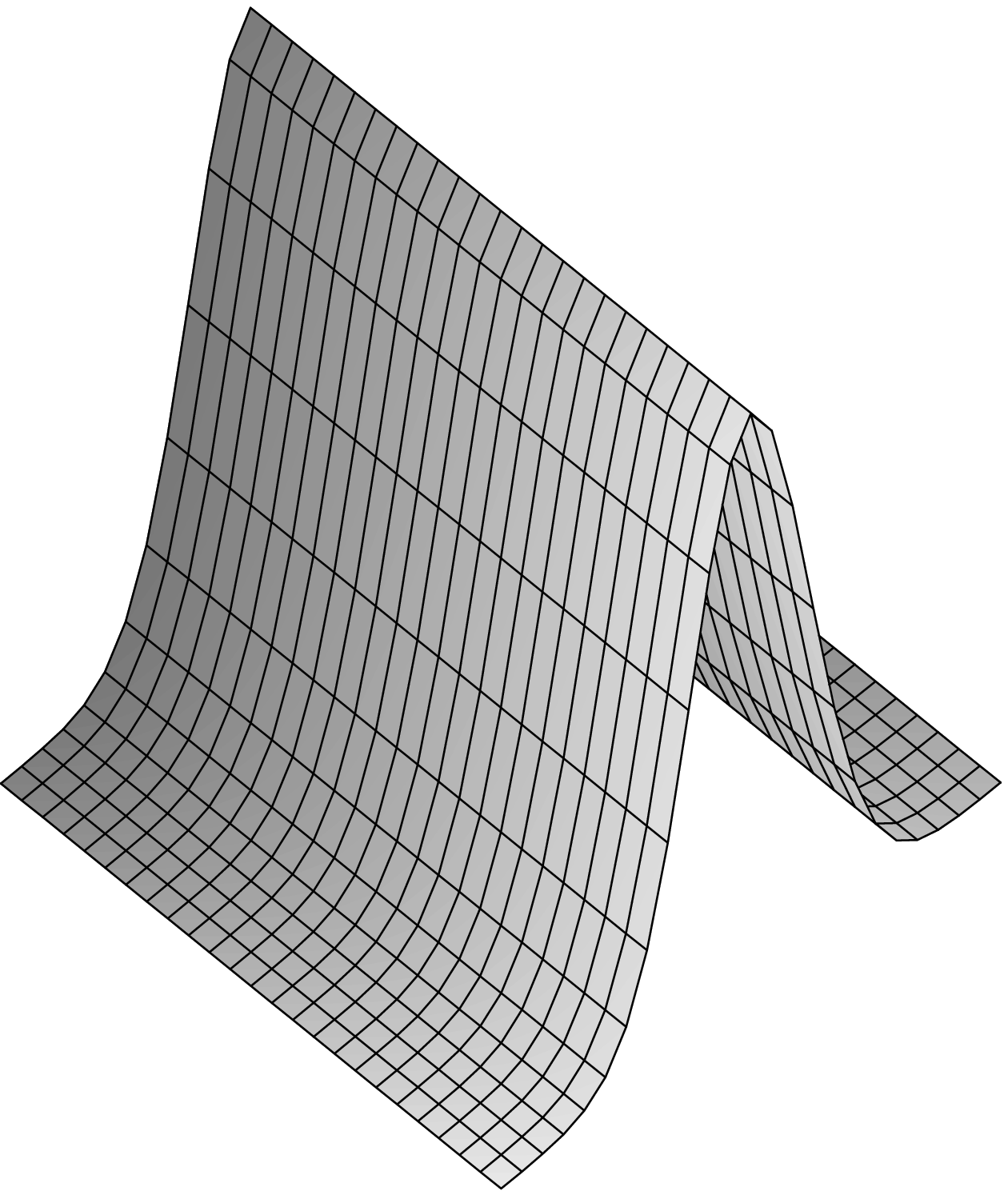}
\end{center}
where the direction of gauge freedom is obvious.

The reason we could add $(d\phi,d\phi,d\phi)$ to the connection $A=(A_1,A_2,A_3)$ without affecting the action is that the connection $(d\phi,d\phi,d\phi)$ is {\bf flat}: its holonomy around each plaquette is zero.  A little thought reveals a simple way of constructing such flat connections.  First, for each directed edge $e$ in the lattice, let $s(e),t(e)\in V$ denote the {\bf source} and {\bf target} (or `tail' and `tip') of $e$, respectively:  
\[
\xy
(-15,0)*{\bullet};(15,0)*{\bullet}**\dir{-} ?(.55)*\dir{>}; 
(-15,2.5)*{s(e)}; (0,2.5)*{e}; (15,2.5)*{t(e)} 
\endxy
\]
Then, if we let 
\[
\phi \maps  V\to \R
\]
be any real-valued function of the vertices of spacetime, and define the {\bfm differential of $\phi$} to be
\[
  d\phi \maps  E \to \R
\]
given by 
\[
  d\phi(e)=\phi(t(e)) - \phi(s(e))
\]
the connection $d\phi$ is flat!  To see this, note that computing the holonomy of $d\phi$ around any loop then amounts to just adding and subtracting the same value at each vertex along the way.  For example
taking the holonomy of $d\phi$ along the boundary $\gamma$ of this plaquette:
\[
\xy
(-12,0)*{\xy
   (0,0)*{\includegraphics{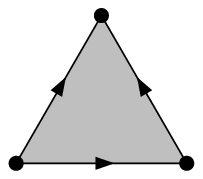}};
   (0,-10)*{e_3};(-6,2)*{e_2};(6,2)*{e_1};            
   (-10,-10)*{v_1};(10,-10)*{v_2};(0,10)*{v_3}; 
               \endxy};
(15,0)*{\displaystyle \gamma=e_1e_2^{-1}e_3}
\endxy
\]
gives
\begin{align*}
  H(d\phi,\gamma) &= d\phi(e_1) - d\phi(e_2) + d\phi(e_3) \\
                           &= (\phi_3 - \phi_2) - (\phi_3 - \phi_1) + (\phi_2 - \phi_1) = 0
\end{align*}
where $\phi_i := \phi(v_i)$ for $i=1,2,3$.  In fact, in lattice gauge theory we define a {\bf gauge transformation} to be a map from the set of vertices to the gauge group.  The space of gauge transformations is thus
\[
           {\cal G} = \R^V.
\]
It is convenient to think of ${\cal G}$ as a group which acts on the space $\cal A$ of connections by
\begin{align*}
         {\cal G}\times {\cal A} &\to {\cal A} \\
               (\phi, A) &\mapsto A + d\phi. 
\end{align*}

A common way of eliminating divergences caused by gauge freedom is {\bf gauge fixing}.  Fixing a gauge means choosing some method to pick one representative of each gauge-equivalence class $[A]$ and doing path integrals over just these.  There are problems with this approach, however.  In a general gauge theory, it might not even be possible to fix a smooth, global gauge over which to integrate.  But even when we can do this, the arbitrary choice involved in fixing a gauge is undesirable, philosophically.  Gauge fixing amounts to pretending a {\em quotient space} of $\cal A$, the space of  {\bf connections modulo gauge transformations}, is a {\em subspace}.  A better approach is to use the quotient space directly.  Namely, modding out by the action of ${\cal G}$ on $\cal A$ gives the quotient space $\cal A /G$ consisting of gauge-equivalence classes of connections, and we do path integrals like
\[
  Z =    \int_{{\cal A}/{\cal G}} e^{-S([A])} d[A].
\]
When gauge fixing works, path integrals over $\cal A/G$ give the same results as gauge fixing.  But integrating over $\cal A/G$ is more general and involves no arbitrary choices.

Since gauge equivalent connections are regarded as physically equivalent, the quotient space ${{\cal A}/{\cal G}}$ of connections mod gauge transformations is sometimes called the {\bf physical configuration space} for vacuum electromagnetism.   A {\bf physical observable} is then any real-valued function on the physical configuration space:
\[
     O: {{\cal A}/{\cal G}} \to \R
\] 
or equivalently, any gauge-invariant function on the space $\cal A$ of connections.

In the case of noncompact gauge group even factoring out all of the gauge freedom may be insufficient to regularize our path integrals.  In particular, there are certain topological conditions our spacetime must meet for this programme to give convergent path integrals.  As a consequence, in the case where the gauge group is $\R$, we must sometimes take more drastic measures in order to extract meaningful results, and this leads to some interesting differences between the cases $G=\R$ and $G=U(1)$ as the gauge group for electromagnetism.  These differences are related to the famous `Bohm-Aharonov effect', which has higher dimensional generalizations in the case of $p$-form electromagnetism. We discuss this issue in Section \ref{sec:cohomology}.

\section{{\boldmath Spacetime Lattice as an $n$-Complex}}

\label{sec:ngraph}

Just as the concept of a manifold is fundamental to an elegant treatment of general relativity,  lattice gauge theory must give due attention to the spacetime lattice itself.  In particular, we should say with some precision what we mean when we describe discrete spacetime as an `$n$-graph.'  We make no attempt in this section at giving the most general definition of $n$-graph suitable for describing arbitrary discrete gauge theories.  Rather, we show that for {\em abelian} gauge theory,  we really only need a weaker notion --- what we shall call an `$n$-complex'.  

{\boldmath
\subsection{The $n$-Graph Problem}
}

Formulating a precise definition of $n$-graph is not an easy task.  Ideally, we should be able to describe $n$-graphs in a purely combinatorial way:  An $n$-graph should consist of sets of various kinds of cells, together with maps telling how the cells are linked together.  An ordinary (directed) {\bf graph}, for example, is specified by a set $V$ of vertices and a set $E$ of edges, together with {\bf source} and {\bf target} maps
\[
      s,t\maps E \to V
\] 
telling at which vertex each edge begins and ends.

If one tries copying this definition to define a notion of $n$-graph, one quickly realizes the immensity of the combinatorial problem at hand!  Even if we only wish to add 2-dimensional faces, complete generality requires an infinite number of new sets and set maps; whereas an edge has only two ends, a 2d face can have arbitrarily many sides:  
\begin{center}
\includegraphics{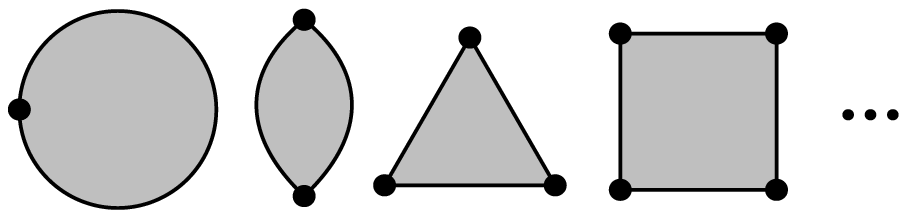}
\end{center}
What's more, the specified maps must satisfy certain equations.  For example, in the pentagonal face
\begin{center}
\includegraphics[width=1.7in]{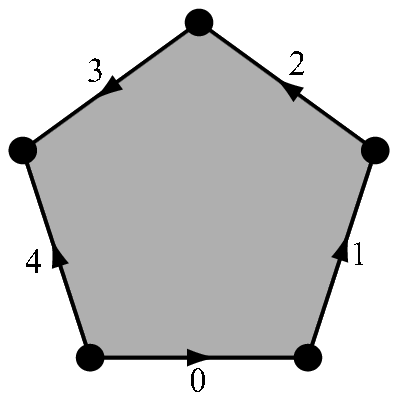}
\end{center}
we must impose equations saying that the $0$th and $4$th edges have the same source vertex, that the target of the $2$nd edge is the source of the $3$rd, and so on (three more equations).   These equations are called {\bf incidence relations}.

And it gets worse.  A 3-cell may have arbitrarily many faces, each of which may have any number of sides \ldots

There are various attempts at formalizing the notion of $n$-graph in a fully satisfactory way, most relying heavily on category theory, or rather $n$-category theory, which itself is still under development.   At the present state of the art, however, we are forced to make compromises of one sort or another.  Each of these compromises carries with it its own weaknesses.  

We could, for example, restrict the allowable shapes of cells.  Allowing only simplices as cells, for example, lets us define `simplicial $n$-graphs' as  simplicial sets \cite{May}, where we simply remove any cells of dimension greater than $n$.  Similarly, we could define `cubical $n$-graphs' using cubical sets \cite{GL}.  If we are only interested in lattice gauge theory as a computational tool, these should be sufficient, and indeed the study of lattice gauge theory began with regular cubical lattices.  However, if we really believe spacetime is discrete at some scale, then there seems to be no {\it a priori} physical justification for imposing such strong conditions on the shapes of cells!   

For this reason, piecewise-linear CW-complexes (or PLCW-complexes, for short) \cite{HMS, H} have been used as discrete models of spacetime in loop quantum gravity \cite{Baez,Baez2}.  These allow a wider variety of cell shapes, and are quite handy if, for example, we want to chop up a manifold with boundary into polyhedra: 
\begin{center}
 \includegraphics
{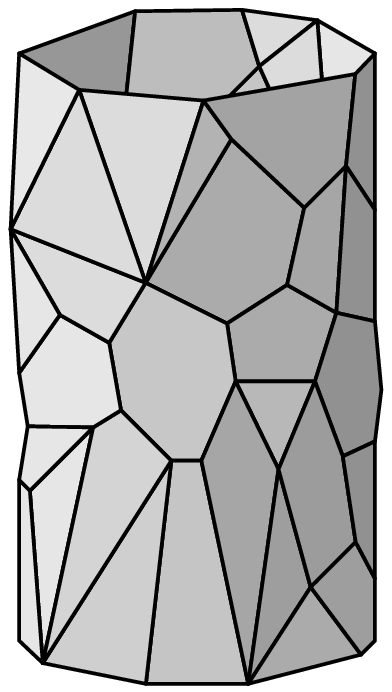}
\end{center}
Unlike simplicial sets, however, we do not know how to define PLCW complexes {\em intrinsically} --- they rely on an ambient space for their construction.   This too is undesirable for physics, analogous to defining the spacetime manifold in relativity as a submanifold of some higher dimensional Euclidean space.  In addition, the polyhedral cells in PLCW complexes would not allow, for example, plaquettes bounded by only one or two edges:
\begin{center}
  \includegraphics{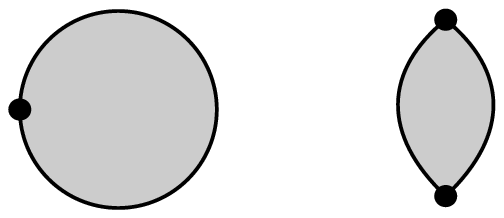}
\end{center}
 or 3-cells bounded by fewer than four plaquettes, like these:  
 \begin{center}
   \includegraphics{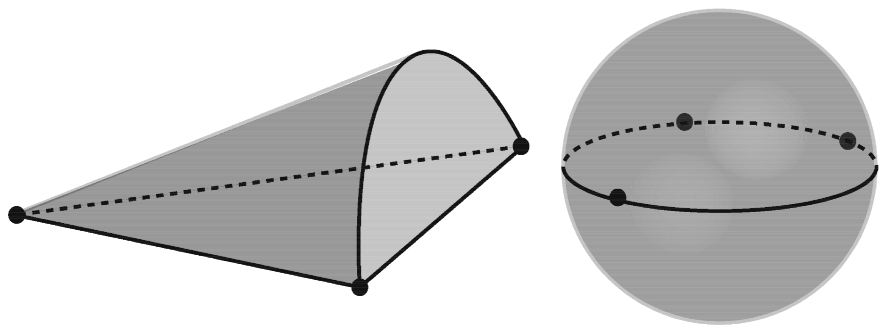}
 \end{center}
 and so on. 
The problem with all of these is that they involve curved sides; cells in PLCW complexes must be constructible in Euclidean space using only polyhedra with straight sides. So in fact, even our simple two-plaquette example from Section \ref{sec:survey} is not a PLCW complex.  Worse, a typical {\em graph} is not a 1-dimensional PLCW complex, so PLCW complexes are hardly a candidate for general $n$-graphs.

{\boldmath
\subsection{$n$-Complexes}
}
\label{ncomplex}
Luckily, it turns out that in gauge theory, and even `higher gauge theory', we can avoid many of the technical issues with $n$-graphs when the gauge group is {\em abelian} --- that is, in the case of $p$-form electromagnetism!  To see why, note that taking the holonomy of a connection along some path through a graph, we multiply group elements of the edges composing the path {\em in order}.  In a nonabelian gauge theory, this requires we keep close track of how the edges are interconnected.  This is manageable for 1-dimensional holonomies, but in `$p$-form gauge theory', where holonomy  involves multiplying group elements labelling $p$-cells, it is difficult to know what order to use.  In general, the holonomy depends intimately on the incidence relations of the $n$-graph --- that is, on the details of how cells are linked up --- and keeping track of this requires a fairly sophisticated notion of $n$-graph.  If the gauge group is abelian, however, then the order in which we apply the group operation is unimportant, and all that matters is the orientation of each cell! 

The key to defining a rudimentary $n$-graph suitable to our purposes is to follow the topological maxim, {\em `The boundary of a boundary is zero'}.  This principle, so phrased by John Archibald Wheeler \cite{CW,MTW}, has limitless applications to physics, from Kirchoff's laws in electrical circuits to the Bianchi identities in general relativity.  It is also the topological foundation of Maxwell's equations themselves, so it is perhaps not so surprising that it plays a foundational role in a precise formulation of lattice electromagnetism.

For us, spacetime will be a \textbf{cell complex} $M$, which we define as follows.  First of all, $M$ consists of a list of sets 
\[
  X_0, X_1, X_2 \ldots , X_k, \ldots
\]
where we call $X_k=X_k(M)$ the set of {\bfm $k$-cells} in $M$.   Now intuitively, the boundary of a $k$-cell should be a ``sum'' of $(k-1)$-cells.  To formalize such sums, for each $k$ we let $C_k = C_k(M)$ be the free abelian group on the set $X_k$.   In other words, $C_k$ just consists of all formal linear combinations of $k$-cells, with integer coefficients.   We call the elements of $C_k$  the {\bfm $k$-chains} in $M$. We then hypothesize \textbf{boundary maps} $\partial_k\maps C_k\rightarrow C_{k-1}$ and require that these be linear over $\Z$.  If we let $\partial_0\maps C_0\to 0$ be the unique map from $C_0$ to the trivial group, then the boundary maps fit together like this:
\[
0 \buildrel \partial_{0} \over \longleftarrow
C_{0} \buildrel \partial_{1} \over \longleftarrow
C_{1} \buildrel \partial_{2} \over \longleftarrow
C_{2} \buildrel \partial_{3} \over \longleftarrow 
\cdots\buildrel \partial_{k} \over \longleftarrow
C_{k} \buildrel \partial_{k+1} \over \longleftarrow 
\cdots.
\]

We will say that the cell complex is {\bfm $n$-dimensional}, or that it is an {\bfm $n$-complex}, provided 
$X_k =\emptyset$ for all $k>n$.\footnote{Note that $X_n$ can be empty.  Thus an $n$-complex is trivially also an $(n+1)$-complex. }
When the cell complex is $n$-dimensional, we might as well truncate the chain and write simply
\[
0 \buildrel \partial_{0} \over \longleftarrow
C_{0} \buildrel \partial_{1} \over \longleftarrow
C_{1} \buildrel \partial_{2} \over \longleftarrow
C_{2} \buildrel \partial_{3} \over \longleftarrow 
\cdots\buildrel \partial_{n} \over \longleftarrow
C_{n}.
\]

With this construction, Wheeler's principle can be stated in a concise way:
\[
   \partial_{k-1}\partial_{k}c=0,
\]
for every $c\in C_k$.  Better yet, when it is clear which boundary maps we are talking about, we often drop the subscripts and just write $\partial$ for all of them.  If we also drop the explicit reference to a $k$-chain $c$, we can write simply
\[
   \partial\partial=0
\]
or even 
\[
   \partial^2=0.
\]
So we conclude our definition of a cell complex by demanding that the boundary of the boundary of any $k$-chain is zero, in the sense just described.

In homological algebra, such a sequence of maps $\partial\maps C_{k} \to C_{k-1}$ satisfying $\partial\partial=0$ is called a {\bf chain complex}.  This lets us summarize the definition in a concise way:

\begin{defn}  An {\bfm $n$-dimensional cell complex}, or {\bfm $n$-complex} $M$ is a chain complex
\label{def:ncomplex}
\[
0 \buildrel \partial \over \longleftarrow
C_{0} \buildrel \partial \over \longleftarrow
C_{1} \buildrel \partial \over \longleftarrow
C_{2} \buildrel \partial \over \longleftarrow 
\cdots\buildrel \partial \over \longleftarrow
C_{n}
\]
 of free abelian groups $C_k(M)$, each equipped with a preferred basis $X_k(M)$.
\end{defn}

In a certain sense, this definition is perhaps too inclusive.  We could take, for instance, the sets $X_k$ of cells to be any sets whatsoever and let each boundary map $\partial$ be the zero map. One can also concoct other sorts of perverse examples, all falling 
squarely within the bounds of the definition, but with seemingly no connection between the boundary maps and the geometric notion of boundary.  

The point is that, although the result might be strange or impractical, one actually can write down a theory of electromagnetism on such $n$-complexes.  Our attitude will be simply that it is up to the user of the definition to decide what additional conditions to impose to guarantee more `reasonable' $n$-complexes.
But to see that our definition of an $n$-complex at least does include the sort of discrete spacetimes we really {\em are} interested in, and to show how the definition might be used in practice, an example or two might be in order.  

\vspace{1em}

\noindent\textbf{Example:} In Section \ref{sec:survey}, we had the following spacetime:
\[
\usebox{\twoface}
\]
But now, instead of simply talking about the sets $X_0=V$, $X_1=E$, and $X_2=P$ of vertices, edges and plaquettes, we make them into free abelian groups
\[
C_0\iso \Z^{V},\, 
C_1\iso \Z^{E},\, \text{ and }
C_2\iso \Z^{P},
\] 
with ordered bases $V$, $E$, $P$, respectively.  The boundary map $\partial \maps C_1 \to C_0$ we define by assigning to each edge $e\in C_1$ its target minus its source:
\[
     \partial(e)= t(e)-s(e).
\]
In particular, since all three edges in this example go from $v_1$ to $v_2$, we have
\begin{align*}
    &\partial(e_1)=v_2-v_1 \\
    &\partial(e_2)=v_2-v_1 \\
    &\partial(e_3)=v_2-v_1.
\end{align*}
Similarly, we define the boundary of a plaquette $p\in C_2$ by taking the sum of the edges around its physical boundary, but with opposite sign if an edge points opposite the plaquette orientation.  Thus, in our example,
\begin{align*}
    &\partial(p_1)=e_1-e_2\\
    &\partial(p_2)=e_2-e_3 .
\end{align*}

It is often convenient to write the boundary maps in matrix form.
Relative to the ordered bases $\{v_1,v_2\}$ of $C_0$, $\{e_1,e_2,e_3\}$ of $C_1$, and $\{p_1,p_2\}$ of $C_2$, they are represented in our example by the matrices:
\[
\partial_{1}=\left[
\begin{array}{rrr}
 -1&-1&-1\\
 1 & 1& 1
\end{array}\right]
\qquad 
\partial_{2}=\left[
\begin{array}{rr}
1&0\\
-1&1\\
0&-1
\end{array}\right].
\]
Obviously $\partial_1 \partial_2$ is just the $2\times 2$ zero matrix, as required.

\vspace{1em}

\noindent\textbf{Example:} As a second example, take the solid tetrahedron:
\[ 
\xy
(1,3)*{\includegraphics{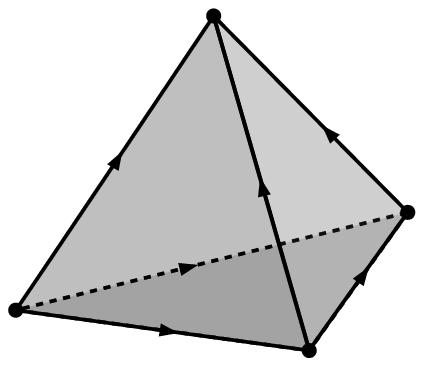}};
(-20,-10)*{}="v1";"v1"+(-2,-1)*{v_1};
(10,-14)*{}="v2";"v2"+(-1,-2)*{v_2};
(20,0)*{}="v3";"v3"+(3,0)*{v_3};
(0,20)*{}="v4";"v4"+(0,3)*{v_4};
(-5,-12) + (0,-3) *{e_1};
(0,-5) + (-1,2) *{e_2};
(-10,5) + (-2,1) *{e_4};
(15,-7) + (1.7,-1.7) *{e_3};
(5,3) + (2,1) *{e_5};
(10,10)+(1,2) *{e_6};
\endxy
\]
which has four vertices, six edges, four plaquettes, and one 3-cell. Here we have left out the labels on plaquettes, to avoid cluttering the diagram, but for convenience we take the $i$th plaquette $p_i$ to be the triangular face whose vertices include all vertices but $v_i$. On each plaquette, let us take the orientation such that its three edges are traversed in numerically increasing order (or a cyclic permutation).  To specify the orientation of the 3-cell $c$, we call each of its faces positively oriented if its orientation is counterclockwise when viewed from inside $c$.  The resulting chain complex then looks like this:
\[
0 \buildrel \partial_{0} \over \longleftarrow
\Z^4 \buildrel \partial_{1} \over \longleftarrow
\Z^6 \buildrel \partial_{2} \over \longleftarrow
\Z^4\buildrel \partial_{3} \over \longleftarrow
\Z
\]
with the boundary maps represented by the matrices
\[
\begin{array}{cc}
\partial_0=
\begin{bmatrix} 0&0&0&0\end{bmatrix} 
&
\partial_1=
\left[ \begin{array}{rrrrrr}
-1& -1& 0& -1& 0& 0\\
1& 0& -1& 0& -1& 0\\
0& 1& 1& 0&0 &-1 \\
0& 0& 0& 1& 1& 1
\end{array}\right]
\\
\partial_2=
\left[ \begin{array}{rrrr}
 0& 0&-1 &-1  \\
 0& -1&0 &1  \\
 -1 &0 &0 &-1  \\
 0 &1 &1 &0  \\
 1 & 0& -1& 0 \\
 -1& -1& 0& 0   
 \end{array}\right] 
 &
\partial_3 = \left[\begin{array}{r} 1\\-1\\1\\-1 \end{array}\right]
\end{array}
\]
Again, we can easily verify that $\partial_0\partial_1$, $\partial_1\partial_2$, and $\partial_2\partial_3$ are all zero, so we really do get a chain complex.

\vspace{1em}

Notice that in passing from the physical model of a spacetime lattice to its corresponding $n$-complex, one is actually throwing away quite a lot of information! Given a chain complex, it may be impossible to reconstruct every detail of how cells in the original model might have been hooked together.  Applying the boundary map to a $k$-cell only tells us `how many times' each $(k-1)$-cell is included in its boundary, with possible cancellation.  For example, in the cylinder with one plaquette $p$:
\[
\xy
   (0,0)*{\includegraphics[height=.2\textwidth] {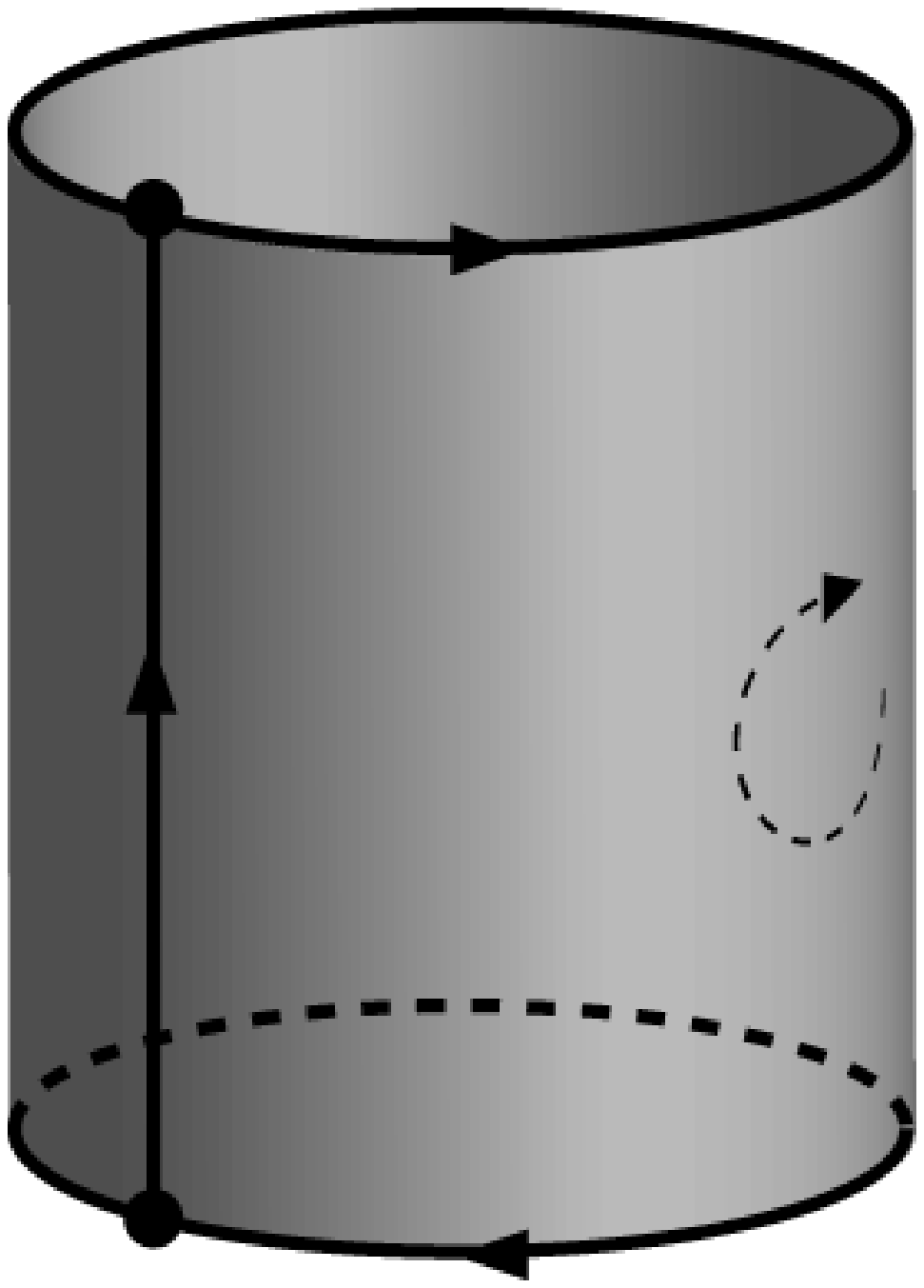}};
   (0,8)*{e_1};
       (0,-18)*{e_3};
       (-5.5,0)*{e_2};
\endxy
\]
we have $\partial p=e_1-e_2+e_3+e_2=e_1+e_3$, leaving no trace of the fact that $e_2$ is actually included twice in the physical boundary of $p$, with opposite orientation.  

We must emphasize that we are free to discard all of these incidence data only because electromagnetism is abelian.  As we shall soon see, almost everything we need for electromagnetism is contained in the  chain complex of spacetime.  We can even think of discrete spacetime as {\em being} its chain complex, a perspective which we explore more fully in Section \ref{sec:chain}.

One immediate generalization of the notion of cell complex is sometimes useful. Although {\em integer} linear combinations of cells are perhaps the most sensible from a purely combinatorial perspective, we may instead take {\em real} coefficients.   The $C_k$ then become real vector spaces, rather than mere abelian groups, allowing arbitrary real-linear combinations of cells in the spacetime lattice.  This is useful because it allows us to exploit the nice properties of the field $\R$ and vector spaces over it in certain computations.  More generally, we could take coefficients in any ring $R$.  Formal $R$-linear combinations of elements of $X_k$ give the {\bfm free $R$-module} on $X_k$.  When we need to be specific about what ring we are using, we write
\[
   C_k(M;R)
\] 
for the $R$-module of $k$-chains in $M$.

\section{Discrete {\boldmath $p$}-Connections and the Action }
\label{sec:action}

\subsection{Cochains as Forms}
\label{sec:cochain}

The local theory of continuum electromagnetism is most beautifully written in the language of differential forms  \cite{GFKG}.
In $p$-form electromagnetism, the analog of the connection $A$ is locally a $p$-form, making the electromagnetic field $F$=$dA$ a $(p+1)$-form.  In {\em lattice} $p$-form electromagnetism, we thus define a {\bfm $p$-connection} to be a map assigning to each $p$-cell an element of the gauge group:
\[
   A\maps X_p(M) \to \R.
\]
But in the framework introduced in the previous section, we can now speak not only of $p$-cells but also formal linear combinations of them.  The real beauty of this viewpoint is we can extend $A$ linearly to obtain a linear functional
\[
   A\maps C_p(M;\R) \to \R.
\]
That is, we can consider a $p$-connection to be an element of the dual space
\[
    C^p(M;\R):=C_p(M;\R)^\ast,
\]
which we dub the space of {\bfm $p$-cochains}.  Likewise, the curvature is a linear functional on the space of $(p+1)$-cells, or an element of $C^{p+1}(M;\R)$.  In general, we define
\[
    C^k(M;R):=
    C_k(M;R)^\ast
\]  
to be the $R$-module dual to $C_k(M;R)$ --- i.e.\ the module of all maps $C_k(M;R)\to R$ that are linear over $R$.  When the ring $R$ is clear from the context (typically $\R$ or $\Z$), we may simply write $C^k$ for $C^k(M;R)$.

To make our discrete theory look as much as possible like its continuous counterpart, we need to understand what takes the place of the differential.  This construction is standard in algebraic topology \cite{GH,Hatcher}.  Since the boundary maps form a chain complex, 
\[
0 \buildrel \partial_{0} \over \longleftarrow
C_{0} \buildrel \partial_{1} \over \longleftarrow
C_{1} \buildrel \partial_{2} \over \longleftarrow
C_{2} \buildrel \partial_{3} \over \longleftarrow 
\cdots\buildrel \partial_{n} \over \longleftarrow
C_{n}
\]
we get a chain complex of the dual spaces in the opposite direction, the {\bf cochain complex}, by taking the adjoint of each boundary map:
\[
C^{0} \buildrel d_{0} \over \longrightarrow
C^{1} \buildrel d_{1} \over \longrightarrow
C^{2} \buildrel d_{2} \over \longrightarrow
\cdots\buildrel d_{n-1} \over \longrightarrow
C^{n} \buildrel d_{n} \over \longrightarrow
0.
\]
That is, given $\omega\in C^i$, we define $d_i \omega\in C^{i+1}$ by
\(
   d_i \omega(x) = \omega(\partial_{i+1} x).
\) 
Following the language of differential forms, we call the maps $d_i$ {\bf differentials}.  Again, dropping the subscripts on both the boundary maps and the differentials usually causes no confusion and we can write more cleanly 
\[
df(x)=f(\partial x).
\]
This definition justifies the name cochain complex: it is easy to see that $\partial\partial=0$ implies $dd=0$.

Exploiting the strong parallel between our cochain complex and the De\-Rham complex of differential forms, we will slip between the two vocabularies at will, calling cochains `forms' whenever convenient.  Also borrowing from differential geometry, we will say that a $k$-cochain $\omega$ is \textbf{closed} if $d\omega=0$ or \textbf{exact} if $\omega=d\eta$ for some $(k-1)$-cochain $\eta$.  The equation $dd=0$ then has its usual De\-Rham translation, ``all exact forms are closed."

Differential forms live to be integrated, so we should understand what it means to integrate a {\em discrete} differential form.  There is a beautiful way of thinking about integrating a cochain that will clarify the correspondence between the equation \(df(x)=f(\partial x)\) and the continuum theory.  The idea is  to think of `integrating' a $k$-cochain $\omega$ over a `region' (a $k$-chain) $R$ as evaluating $\omega(R)$.  If we introduce the notation
\[
    \int_R \omega :=\omega(R),
\]
then  the definition of the differential $d$ as the adjoint of the boundary map $\partial$ is just
\begin{equation*}
   \int_{R} d\omega = \int_{\partial R} \omega
\end{equation*}
--- a discrete version of Stokes' theorem on differential forms!

\subsection{Action}
\label{action}
In ordinary continuum electromagnetism, the Lagrangian for the free electromagnetic field $F=dA$ is given by\footnote{In many books, one sees the free electromagnetic Lagrangian written as ${1\over 2}F\wedge\star F$, or in index notation as $\fourth F_{\mu\nu} F^{\mu\nu}$, without the coupling constant $(1/e^2)$.   This means either that the authors have hidden the coupling constant elsewhere, usually by defining the covariant derivative to be $D_\mu = \partial_\mu - ieA_\mu$, or that they are interested only in the {\em classical} theory, where constant multiples of the action have no effect. After quantization, the magnitude of $e$ scales the probability of large vacuum fluctuations.
}  
\[
   {\cal L}=\frac{1}{2e^2} F\wedge\star F.
\]
This equation holds for free $p$-form electromagnetism as well, where $F$ is the $p$-form field $F=dA$. 
To write down a lattice version of this theory, we thus seem to need discrete analogs of the wedge product $\wedge$ and the Hodge star operator $\star$.  The Hodge star in differential geometry turns $q$-forms into $(n-q)$-forms, so we seem to need a map:
\[
    \star \maps  C^q \to C^{n-q}
\]
for each $q=0, 1, \ldots, n$.  Likewise, for an analog of the wedge product of differential forms we would want
\[
    \wedge \maps  C^q \times C^r \to C^{q+r}.
\]

Defining these maps seems to be the main obstacle to completely formulating a discrete analog of the theory of differential forms.   In certain special cases, it is known what these operations should look like.  For example, for simplicial complexes, the best approximation to the wedge product of differential forms is the `cup product' $\cup$ of cochains \cite{GH}.   For the Hodge star, a cubical complex is more convenient, since Hodge duality is related to Poincar\'e duality, and the Poincar\'e dual of a cubical lattice is also cubical.\footnote{For more on discrete differential geometry, we refer the reader to the papers by Forgy and Schreiber \cite{FS} and Forman \cite{Forman}, and to additional references which they cite.}

But we would rather not impose any restrictions on the shapes of cells in our spacetime model.  Forcing the lattice to be too regular goes against relativity's lesson that spacetime, at least on the macroscopic level, does not have preferred directions the way a crystal lattice does.     

Luckily, if we resist the temptation to dissect the formula for the action
\[
   S=\frac{1}{2e^2}\int F\wedge\star F,
\]
we can instead discretize the whole thing in one fell swoop!  The key observation is that
\[
   \langle F,G\rangle := \int F\wedge\star G
\]
defines an inner product, the {\bf Hodge inner product}, on differential $(p+1)$-forms.  So in the lattice theory, if we endow $C^{p+1}$ with an inner product 
\[
\langle\phantom{F},\phantom{F}\rangle \maps  C^{p+1}\times C^{p+1}\to \R,
\] 
then we can define the action corresponding to the field $F$ to be 
\[
    S=\frac{1}{2e^2}\langle F,F \rangle.
\]

What this inner product should look like for a given lattice depends on the structure and the dimension of the lattice.  We shall not dig too deeply into this issue, but just mention a few `reasonable' requirements. For this it is useful to write the inner product as:
\[
      \langle F,G \rangle = h^{ij}F(x_i) G(x_j)
\]
where the implicit sum is over all pairs of $(p+1)$-cells $x_i, x_j$. 
One sensible condition that might be imposed on this inner product is a `locality condition.'  That is, we do not expect the action to have correlations between distant points in spacetime, so we might require that $h^{ij}$ vanish whenever $x_i,y_i$ are `nonadjacent' $(p+1)$-cells.  Also, when the details of lattice geometry are carefully accounted for, we might expect the correlation between two cells to depend on the physical {\em angle} between those cells.   In particular, on a regular cubical lattice, we might well expect $h^{ij}$ to be a diagonal matrix.  Note that neither the locality condition nor any geometric data used to build the inner product can be derived from the chain complex itself.  Both constitute additional structure in the spactime model.  

For the sake of simplicity, many of our examples in this paper involve electromagnetism in two dimensions.  For this reason, we next turn briefly away from our more general development of the theory to consider this special case.  More generally, we consider the case of $p$-form electromagnetism in $p+1$ dimensions and find that the inner product on $p$-cochains has a natural and simple form.
{\boldmath
\subsection{The Case of {$p+1$} Dimensions}
}
\label{pplus1}
Ordinary gauge theories in {\em two} dimensions are famously well-behaved \cite{Witten}.  One of the reasons electromagnetism in two dimension is especially nice is that the electromagnetic field, being a 2-form, can be written simply as a scalar field times the volume form (or perhaps more properly, the `area form'):
\[
              F= f \; \vol.
\]
This allows us to choose the best possible inner product on $1$-cochains --- an exact analog of the Hodge inner product --- without having to define a discrete Hodge star operator in detail.  In particular, the only Hodge duals we really need to consider are the trivial ones: $\star \vol = 1$ and $\star 1 = \vol$.  More generally, and for the same reason, we can do discrete $p$-form electromagentism and avoid any unpleasant entanglements with the Hodge star as long as we work in $(p+1)$-dimensional spacetime.

To do path integrals, there are several reasonable choices we could make for the action.  We would like, however, to arrive at a discrete theory that looks as much as possible like the more familiar theory of electromagnetism on a manifold.  With this motivation, we make an unapologetic appeal to the known continuum theory to decide which action to use. 

Since the usual Lagrangian for vacuum electromagnetism is $L=\frac{1}{2e^2}F\wedge \star F$, we find, in the two-dimensional case:
\begin{align*}
          L&=\frac{1}{2e^2}(f \; \vol)\wedge\star(f \; \vol) \\
           &=\frac{1}{2e^2}f^2 \;\vol\wedge 1 \\
           &=\frac{f^2}{2e^2} \vol.
\end{align*}

Now, if we want to discretize this theory, say for the purpose of modeling on a computer, we clearly want a fine enough lattice that the value of the scalar curvature field $f$ in the above equations is essentially constant over any plaquette.  If we make this assumption, then the action associated with a plaquette $p_i$ whose curvature is $F$ should be
\begin{align*}  
    S(p_i)&= \int_{p_i} L \\
          &= \int_{p_i} \frac{f^2}{2e^2} \;\vol \\
          &\approx \frac{f^2}{2e^2}\int_{p_i}\vol \\
          &= \frac{f^2}{2e^2}\cdot V_i.
\end{align*}
where $V_i$ is the area of the plaquette $p_i$.  Similarly, the most obvious way of discretizing the curvature is to assign to each plaquette the area integral of its continuum curvature.
\begin{align*}
      F_i &= \int_{p_i} F \\
          &= \int_{p_i} f \;\vol \\
          &\approx f\int_{p_i}\vol \\
          &= f\cdot V_i.
\end{align*}

Combining these last two results we find a candidate for the action which now uses only variables that are available to us directly in the lattice model!
\[
        S_i = \frac{1}{2e^2} \frac{F_i^2}{V_i}
\]
To get the total action, we just sum over plaquettes.  Recalling from Section \ref{action} that the action is given by
\[
    S=\frac{1}{2e^2}\langle F,F \rangle,
\]
the corresponding inner product on $2$-cochains for electromagnetism in 2 dimensions is thus
\[
     \langle F,G \rangle = \sum_{p_i\in X_2} \frac{F_i G_i}{V_i}.
\]

The preceding derivation of the action did not rely explicitly on spacetime being two dimensional but only on the fact that the curvature is a scalar multiple of the volume form.  More generally, this happens for $p$-form electromagnetism whenever spacetime is $(p+1)$-dimensional.  Indeed, in this case we can repeat the above calculations to obtain the obvious generalization of the action.  The inner product on $(p+1)$-cochains for $p$-form electromagnetism in $p+1$ dimensions is thus 
\[
     \langle F,G \rangle = \sum_{p_i\in X_{p+1}} \frac{F_i G_i}{V_i}.
\]
Here $V_i:=\vol(x_i)$, where  
we now use $\vol$ to denote the {\bf discrete volume form} --- a $(p+1)$-cochain 
\[
           \vol\maps C_{p+1}\to \R
\]
that assigns to each $(p+1)$-cell its volume.  

\section{Discrete {\boldmath $p$}-form Maxwell Equations}
\label{sec:maxwell}
As with any field theory, we may obtain the classical equations of motion --- in this case the discrete analog of Maxwell's equations, or rather their $p$-form generalization --- by extremizing the action.  Formally, the action in our theory is the map
\[
   S\maps  C^{p}\to \R
\]
given by 
\[
        S(A) = \frac{1}{2e^2}\langle dA,dA\rangle
\]
where $\langle -,-\rangle$ is the inner product on $(p+1)$-cochains introduced in the previous section.   The extrema of this function are the $p$-connections $A$ such that the differential 
\[
 \delta S \maps  T_{A}C^p \to T_{S(A)}\R \iso \R
\]
is the zero map.  We get:
\begin{align*}
0=\delta S &= \frac{1}{2e^2}\delta \langle dA,dA \rangle \\
              &= \frac{1}{e^2} \langle \delta dA,dA \rangle \\
              &= \frac{1}{e^2} \langle d \delta A,dA \rangle \\
              &= \frac{1}{e^2} \langle \delta A,d^\dagger dA \rangle 
\end{align*}
where equality in the second line follows from the product rule together with the symmetry of the inner product, and $d^\dagger$ is the Hilbert space adjoint 
\[
   d^\dagger\maps C^{p+1} \to C^{p}
\]
of the linear operator $d$.  At an extremum, $\delta S = 0$ must hold for all values of the variation $\delta A$.  So, by our above calculation,
\[
    d^{\dagger }dA = 0
\]
or 
\[
    d^{\dagger}F = 0
\]
which is our discrete analog of the vacuum Maxwell equations\footnote{This equation corresponds to the two nontrivial vacuum Maxwell equations, which say $\nabla \cdot \vec E = 0$ and $\nabla \times \vec B = \frac{\partial \vec E}{\partial t} $.  The other two equations come from the tautology $dF=0$, just as with differential forms.}
\[
  \star d {\star}F = 0.
\]

It is worth noting the relationship between the continuum and lattice versions of the Maxwell equations.  In particular, this means understanding the relationship between the two kinds of duality showing up in these equations. 
In Riemannian signature, the Hodge operator satisfies $\star^2 \omega = (-1)^{p(n-p)}\omega$ when acting on any $p$-form $\omega$.  So given a $p$-form $A$ and $(p+1)$-form G, 
we have
\begin{align*}
     \langle dA, G \rangle &= \int dA \wedge \star G \\
                                           &=(-1)^{p+1}\int A \wedge d\star G \\
                                           &=\int A \wedge \star (-1)^{(n-p)p} \star d\star G \\
                                           &=\langle A, (-1)^{(n-p)p} \star d\star G \rangle
\end{align*}
where in the second step we did an integration by parts. Thus
\[
               d^\dagger =  (-1)^{(n-p)p} \star d\star .
\]

In fact, the vacuum $p$-form Maxwell equations:
\[
        d^\dagger F = 0
\]
can be simplified further in the present context.   If $F=dA$ satisfies this equation, then we have:
\[
     0 = \langle d^\dagger F, A \rangle = \langle F, dA \rangle = \langle F, F \rangle.  
\]
In Lorentzian physics, we draw no strong conclusions from this calculation.  But in the Riemannian case,  the inner product $\langle \phantom{F},\phantom{F}\rangle$ is {\em positive definite}, and hence the classical equations of motion for free discrete $p$-form electromagnetism reduce to:
\[
    F=0!
\]
This is shocking if we are accustomed to electromagnetism on Minkowski spacetime.  It means, in particular, that our theory has no analog of electromagnetic waves propagating through a vacuum!  Moreover, the same proof works not only in the discrete case but also, for example, for $p$-form electromagnetism on a compact Riemannian manifold.   The two assumptions leading to the conclusion that $F=0$ are that (1) we have an inner product defined on {\em all} $(p+1)$-forms $F$, and (2) that this inner product is positive definite.  In Minkowski spacetime, the Hodge inner product $\int F\wedge \star G$ is neither defined for all $(p+1)$-forms, nor positive definite, so even compactly supported $p$-form electromagnetic fields in the vacuum need not vanish.

There is one way we might try to weasel out of having such a trivial classical theory in cases where conditions (1) and (2) are met.   Namely, we can take a pre- gauge theory perspective on electromagnetism, and take $F$ as the fundamental field of classical electromagnetism, rather than $A$.  If the electromagnetic field $F$ is not necessarily the differential of any gauge potential, then $dF = 0$ is not automatically satisfied, so we must include it explicitly in the $p$-form Maxwell equations:
\[
     dF=0 \text{ and } d^\dagger F=0.
\]
We will see that these equations {\em do} have solutions other than $F=0$, but only on spacetimes with a certain topological properties --- roughly, spacetimes with `holes' of the type which can be enclosed by a surface of dimension $p+1$.  What we will find is that classical $p$-form electromagnetism in Riemannian signature, in cases where we have an inner product defined on all  $(p+1)$ forms, is what we might call a {\bf topological {\em classical} field theory}:   its space of solutions is determined up to canonical isomorphism by the topology of spacetime!   This, in fact, is true whether we take $A$ or $F$ as the basic field, though the two perspectives disagree on what aspect of the topology determines the space of solutions.

In the next section, we quantize $\R$ $p$-form electromagnetism and, in parallel, develop enough topology to state a criterion for the convergence of the path integral.  As this topology --- cohomology --- is precisely what we need to fully understand the comments above about the classical theory being topological, we return to the $p$-form Maxwell equations in Section \ref{hodge}.

\section{Cohomology and Path Integrals}
\label{sec:cohomology}

\subsection{A Criterion for Path Integral Convergence}
\label{cohomology1}
Consider some small portion of a complex of $\R$-valued cochains 
\[
  \xymatrix{
    C^{q-1} \ar[r]^{d_{q-1}} & C^{q} \ar[r]^{d_{q}}  & C^{q+1}
                 }
\]
Cohomology classifies topological spaces by comparing two subspaces of $C^q$:
\[
\begin{array}{ll}
Z^q:=\ker d_{q} & \text{--- the space of 
$q$-\textbf{cocycles}} \\
B^q:=\ran d_{q-1} & \text{--- the space of 
$q$-\textbf{coboundaries}}
\end{array}
\]
For the cochain complex of {\em any} spacetime lattice we have, by the defining equation $dd=0$,
\[
 B^q\subset Z^q.
\]
That is, every $q$-coboundary is a $q$-cocycle.  Whether the converse of this statement is true depends on the particular topology of the spacetime lattice.  If every $q$-cocycle is a $q$-coboundary, so that $B^q$ and $Z^q$ are {\em equal}, we say that the cochain complex is \textbf{exact} at $C^q$.  In topologically interesting spacetimes (or regions of spacetime), exactness may fail, and we measure the failure of exactness by taking the quotient space
\[
    H^q:=Z^q/B^q,
\]
called the {\bfm $q$th cohomology with real coefficients}. 

To see the ramifications of cohomology in lattice electromagnetism, let us reexamine what caused the path integral 
\[
   Z = \int_{C^p} e^{-S(A)} {\cal D}A 
\]
to diverge in Section \ref{sec:survey}.  In gauge theory, the action $S(A)$ is invariant under gauge transformations.  In our lattice version of $p$-form electromagnetism with gauge group $\R$, gauge transformations take the form
\[
   A \mapsto A + d\varphi,
\]
where $\varphi$ is an arbitrary $(p-1)$-cochain.  In other words, two $p$-connections $A, A'\in C^p$ are {\bf gauge equivalent} if they differ by a $p$-coboundary $d\varphi\in B^p$.  In Section \ref{sec:survey} we saw that the existence of gauge freedom implied path integrals such as the one above must diverge.  Indeed, if ${\cal G}=B^p$ is nontrivial, then we can write
\[
            C^p=B^p \oplus (B^p)^\perp
\]
and express any $p$-connection $A$ uniquely as a sum
\[
           A=A_0 + A^\perp \text{ with } 
  \begin{array}{l} A_0\in B^p \\
                   A^\perp\in(B^p)^\perp
   \end{array}
\]
This lets us break the path integral up as 
\[
   Z = \int_{B^p}\int_{(B^p)^\perp} e^{-S(A_0+A^\perp)} 
                dA^\perp dA_0 .  
\]
But since the action is gauge invariant, $S(A_0+A^\perp)=S(A^\perp)$, and so this becomes
\begin{align*}
  Z &= \int_{B^p}\int_{(B^p)^\perp} e^{-S(A^\perp)} 
                dA^\perp dA_0   \\
           &= \int_{B^p}dA_0 \int_{(B^p)^\perp} e^{-S(A^\perp)} 
                dA^\perp=\infty,
\end{align*}
since for each fixed value of $A^\perp$, the integral over $B^p$ diverges.

To eliminate these divergences it is therefore {\em necessary} that we eliminate any gauge freedom from our space of $p$-connections.  The standard procedure is to pass to the quotient space
\[
\frac{\cal A}{\cal G} = \frac{C^p}{B^p}= \text{\bfm $p$-connections modulo gauge transformations},
\]
effectively declaring {\em gauge equivalent} $p$-connections to be equal.

In many cases, integrating over $C^p/B^p$ instead of just $C^p$ does indeed make the path integral finite. But this does not always work! Whether it works or not depends on a topological condition --- the {\em cohomology} of the lattice.  While factoring out gauge freedom is necessary, it is only sufficient if the $p$th cohomology, $H^p$ is trivial.  Indeed, if
\[
   H^p=\frac{Z^p}{B^p}\neq 0
\]
then since
\[
  \frac{\cal A}{\cal G} = \frac{C^p}{B^p} \iso \frac{C^p}{Z^p} \oplus \frac{Z^p}{B^p}
\]
we have
\begin{align*}
  \int_{{\cal A}/{\cal G}} e^{-S(A)} 
                dA   
 &= \int_{Z^p/B^p}\int_{C^p/Z^p} e^{-S(A)} 
                dA_1 dA_2   \\
 &= \int_{Z^p/B^p}dA_2 \int_{C^p/Z^p} e^{-S(A_1)} 
                dA_1    
           =\infty,
\end{align*}
since $S(A)$ really only depends on $dA=d(A_1 + A_2)=dA_1$.

Pondering more carefully what caused the divergence, we realize that to eliminate all divergences from $\R$ electromagnetism we must use the  quotient space
\[
\frac{\cal A}{{\cal A}_0} = \frac{C^p}{Z^p}= \text{ \bfm $p$-connections modulo flat $p$-connections}
\]
For any lattice with only finitely many $p$-cells, integrating over this space {\em always} gives us a convergent path integral.  The point is that, when $H^p$ is nontrivial, the quadratic form $S$ is still degenerate, even after factoring out all of the gauge freedom.  However, when we factor out not just gauge transformations, or in other words `pure gauge' $p$-connections, but {\em all} flat $p$-connections,  $S$  becomes a positive {\em definite} quadratic form, so that $\int e^{-S}$ is finite.  Visually, on the space of $p$-connections modulo flat $p$-connections, every cross section of the graph of $e^{-S}$ looks just how we expect a well-behaved Gaussian to look:
\begin{center}
  \includegraphics[width=2 in,height=1in]{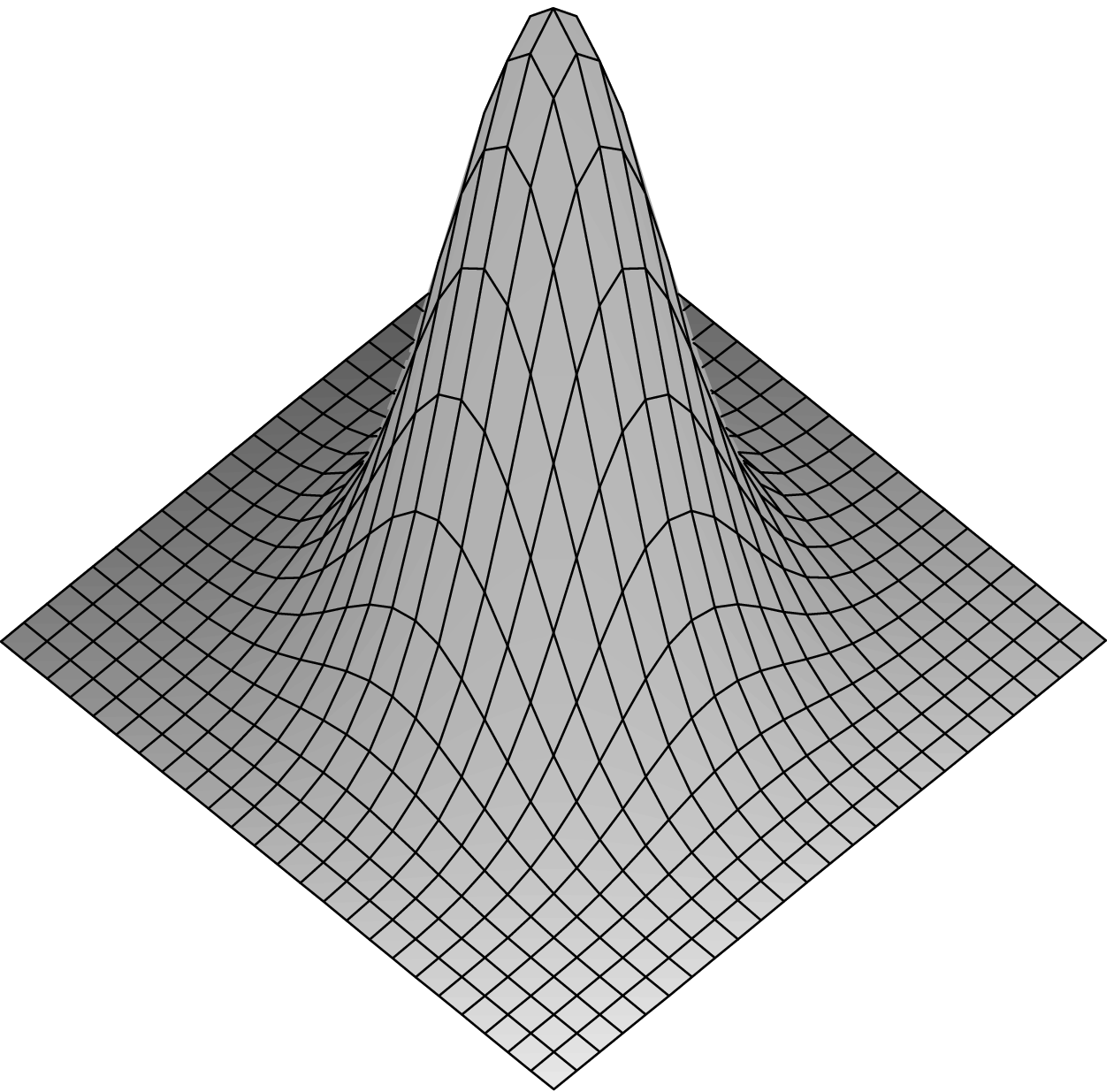}
\end{center}
in contrast with the analogous picture from Section \ref{sec:survey}

\vspace{1em}
Let us consider two examples: one in which modding out by gauge transformations is sufficient to render the path integral finite, and one in which it is not.  First, let us finally fix the problem we found in Section \ref{sec:survey}, namely the divergence of the path integral for 
\[
\usebox{\twoface}
\]
over the space of connections.  We write the connection as
\[
   A = A_1 e^1 + A_2 e^2 + A_3 e^3 = A_i e^i
\]
where $\{e^i\}$ is the basis of 1-cochains dual to the basis $\{e_i\}$ of edges.   A gauge transformation 
\[
   \phi = \phi_1 v^1 + \phi_2 v^2
\]
acts on the connection by $A \mapsto A + d\phi$ where
\[
      d\phi = 
      (\phi_2-\phi_1)(e^1+e^2+e^3).
\]
Since the number $\phi_2-\phi_1 \in \R$ may be chosen arbitrarily, the space of gauge transformations  is just the 1-dimensional subspace
\[
  B^1 := \ran d_0 ={\rm span}\{e^1+e^2+e^3\} \subset C^1
\]

Similarly, the curvature $F$ is given by
\[
  F := dA = 
  (A_1-A_2)p^1-(A_3-A_2)p^2
\]
so the space of flat connections is
\begin{align*}
{\cal A}_0 = Z^1:= \ker d_1&= \{ A\in C_1 | A_1 = A_2 = A_3\} \\
                          &={\rm span}\{e^1+e^2+e^3\} = B^1.
\end{align*}
Thus, due to  the simple topology of the model, the flat connections in this case are precisely the pure gauge connections.  In other words, the first cohomology is trivial:
\[
   H^1 = Z^1/B^1 = \{0\}.
\] 

Now let us do the path integral over the space $C^1/B^1 = C^1/Z^1$.  To do this we make use of the canonical isomorphism
\[
     \frac{C^1}{B^1} \iso (B^1)^\perp
\]
given by associating with each gauge equivalence equivalence class $[A]\in C^1/B^1$ its unique member which is also an element of $(B^1)^\perp$.  
We first define a new orthonormal basis
\begin{align*}
  e^{1'} &= \frac{1}{\sqrt{3}}(e^1 +e^2 + e^3) \\
  e^{2'} &= \frac{1}{\sqrt{2}}(e^1 -e^2) \\
  e^{3'} &= \frac{1}{\sqrt{6}}(e^1 + e^2 - 2 e^3)
\end{align*}
of $C^1$ chosen so that $\{e^{1'}\}$ is a basis for $B^1$ and $\{e^{2'}, e^{3'}\}$ is a basis for $(B^1)^\perp$.   

We can write the connection $A$ in terms of the new dual basis: 
\begin{align*}
  A = A_{i'} e^{i'} 
      &= \frac{A_1 + A_2 + A_3}{\sqrt{3}}e^{1'} + \frac{A_1 -A_2}{\sqrt{2}}e^{2'} +\frac{A_1+A_2-2A_3}{\sqrt{6}}e^{3'} 
\end{align*}  
Then the action becomes
\begin{align*}
     S(A) &=\frac{1}{2e^2}\left(\frac{(A_1 - A_2)^2}{V_1}+\frac{(A_2- A_3)^2}{V_2}\right)= \frac{1}{2e^2} [ (A_1-A_2)^2 + (A_2 - A_3)^2 ] \\
              &= \frac{1}{2e^2 V_1 V_2} \left[ V_2(\sqrt{2} A_{2'})^2 + V_1(-\frac{1}{\sqrt{2}}A_{2'} + \sqrt{\frac{3}{2}}A_{3'})^2 \right]\\
              &=\frac{1}{2e^2 V_1 V_2}
                \begin{bmatrix} A_{2'} & A_{3'} \end{bmatrix}
                \begin{bmatrix} 2V_2 + \frac{1}{2} V_1 & \frac{\sqrt{3}}{2}V_1 \\
                                          \frac{\sqrt{3}}{2}V_1 & \frac{3}{2}V_1 \end{bmatrix}
                \begin{bmatrix} A_{2'} \\ A_{3'} \end{bmatrix} \\
                &=\frac{1}{2e^2V_1 V_2}  \langle A , QA\rangle 
\end{align*}
Now integrating over $\cal{A / G}$ is a snap!  Since the $2\times 2$ matrix in the above expression is nonsingular, the formula we tried to use for this same problem in Section \ref{sec:survey} now gives a finite result:
\begin{align*}
  Z &= \int_{R^2} e^{-S(A)} d^2\!\! A =  
  \sqrt{ \frac{(2\pi e^2 V_1 V_2)^2}{\det( Q)}  } = 2\pi e^2 \sqrt{\frac{V_1 V_2}{3}}. 
\end{align*}  

As a second example, consider again the case of ordinary 1-form electromagnetism in two dimensions, but this time in a spacetime with the topology of a torus:
\[
\xy
(-45.5,0)*{\includegraphics{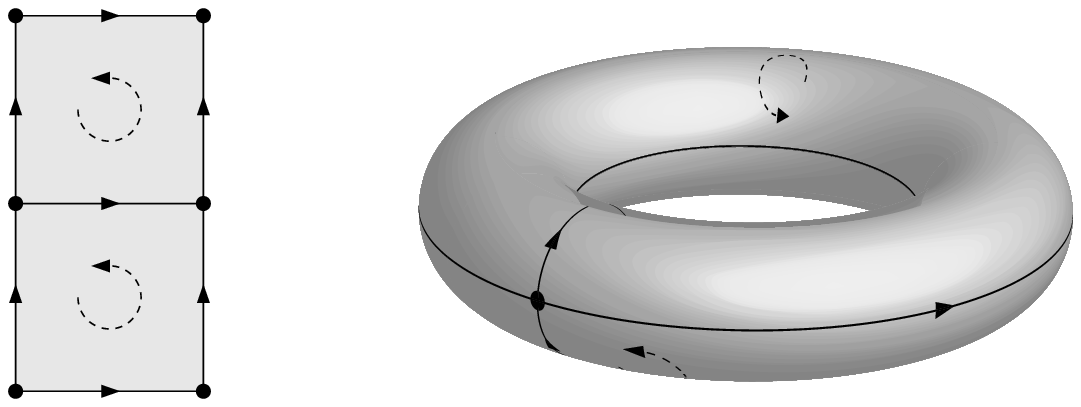}};
(-45.5,0)*{\xy
	(-11,21)*{\displaystyle v_1};
	(11,21)*{\displaystyle v_1};
	(-11,-21)*{\displaystyle v_1};
	(11,-21)*{\displaystyle v_1};
	(-12,0)*{\displaystyle v_2};
	(12,0)*{\displaystyle v_2};
	(0,22)*{\displaystyle e_1};
	(0,-22)*{\displaystyle e_1};
	(-12,-11)*{\displaystyle e_2};
	(12,-11)*{\displaystyle e_2};
	(0,2)*{\displaystyle e_3};
	(-12,11)*{\displaystyle e_4};
	(12,11)*{\displaystyle e_4};
	(0,10)*{\displaystyle p_2};
	(0,-10)*{\displaystyle p_1};
\endxy};
\endxy
\]
Note in the diagram that objects with the same label are identified.  Let us first calculate the cohomology of this lattice.  The cochain complex 
\[
\xymatrix{
     0 \ar[r] &C^0 \ar[r]^{d_0} & C^1 \ar[r]^{d_1} & C^2 \ar[r] & 0}
\]
has differentials represented by the matrices
\[
 d_0 = 
 \left[ \begin{array}{rr} 0&0\\-1&1\\ 0&0\\1&-1\end{array} \right] 
 \qquad
 d_1 =
 \left[ \begin{array}{rrrr} 1&0&-1&0 \\ -1&0&1&0\end{array} \right] 
\]
relative to the bases $\{ v^1, v^2, v^3\}$ of $C^0$, $\{ e^1,e^2,e^3,e^4\}$ of $C^1$, and $\{p^0,p^2\}$ of $C^2$.  We are particularly interested in the first cohomology, since it relates flat connections to gauge transformations.  Some basic linear algebra shows that
\begin{align*}
  Z^1 &:=\ker d_1 = {\rm span}\{e^1+e^3, e^2, e^4\} \iso \R \\
  B^1&:= \ran d_0 = {\rm span}\{e^2 - e^4\} \iso \R
\end{align*}
and therefore
\[
  H^1 = \frac{Z^1}{B^1} \iso \frac{\R^3}{\R} \iso \R^2
\]
Physically, this says there are flat connections $A$ on the torus which are not gauge equivalent to the trivial connection, and that there are two degrees of freedom for such connections.   In electromagnetism we also interpret $H^1 \iso \R^2$ by saying that there are two {\bf Bohm-Aharonov modes} \cite{GFKG, Zee}. 

For the sake of characterizing the topology more fully, we should calculate the other cohomology groups as well.  We find
\[
  H^0 = \frac{{\rm span}\{v^1-v^2\}}{\{0\}} \iso \frac{\R}{\{0\}} \iso \R
\]
and 
\[
  H^2 = \frac{C^2}{{\rm span}\{p^1-p^2\}} \iso \frac{\R^2}{\R} \iso \R.
\]
For doing path integrals on this lattice, it is convenient to take the following orthonormal basis for $C^1$:
\begin{align*}
 e^{1'} &= \frac{1}{\sqrt{2}} (e^1 - e^3) \\
 e^{2'} &= \frac{1}{\sqrt{2}} (e^1 + e^3) \\
 e^{3'} &= \frac{1}{\sqrt{2}} (e^2 + e^4) \\
 e^{4'} &= \frac{1}{\sqrt{2}} (e^2 - e^4) 
\end{align*}
since then $\{e^{1'}\}$ is a basis for $(Z^1)^\perp$ and $\{e^{1'},e^{2'},e^{3'}\}$ is a basis for $(B^1)^\perp$.  In terms of this new basis we can write the connection as
\begin{align*}
   A 
     = A_{i'} e^{i'} &= \frac{(A_1 - A_3)}{\sqrt{2}}e^{1'}
            + \frac{(A_1 + A_3)}{\sqrt{2}}e^{2'} + \frac{(A_2 + A_4)}{\sqrt{2}}e^{3'} 
            + \frac{(A_2 - A_4)}{\sqrt{2}}e^{4'}.
\end{align*}
For simplicity, let us assume that the area of each plaquette is 1, so the action becomes
\begin{align*}
S(A) &= \frac{1}{2e^2} ((A_1-A_3)^2 + (A_3 - A_1)^2) = \frac{2}{e^2} A_{1'}^2. 
\end{align*}
Now just as in the previous example, integrating over connections mod gauge transformations, i.e.\ $C^1/B^1$ is the same as integrating over $(B^1)^\perp= {\rm span}\{e^{1'},e^{2'},e^{3'}\}$, relative to which basis $A$ has components $(A_{1'},A_{2'},A_{3'})$.  But $S(A)$ is  obviously a degenerate quadratic form in the variables $A_{1'},A_{2'},A_{3'}$, so the path integral diverges:
\[
   \int_{C^1/B^1}  e^{-S(A)} {\cal D} A = \int_{\R^3}  e^{-2{A_{1'}}^2/e^2}  dA_{1'}dA_{2'}dA_{3'} = \infty.
\]

Similarly, integrating over connections mod flat connections, i.e.\ $C^1/Z^1$ is the same as integrating over $(Z^1)^\perp$.  But $S(A)$ is obviously nondegenerate in the variable $A_{1'}$, which is the coordinate of $A$ relative to the basis $\{e^{1'}\}$ of $(Z^1)^\perp$.  In fact, in this case we get:
\[
   \int_{C^1/Z^1}  e^{-S(A)} {\cal D} A = \int_\R  e^{-2{A_{1'}}^2/e^2}  dA_{1'} = \sqrt{\frac{\pi}{2}}e.
\]

So we see that to regularize the path integral for the torus, in $\R$ electromagnetism we must kill off both of the Bohm-Aharonov modes, in addition to factoring out gauge freedom. 
We have seen that the criterion in $\R$ electromagnetism for path integrals over the physical configuration space ${\cal A/G}$ to converge is that the first cohomology be trivial:
\[
\left(\begin{array}{c}
  Z = \int_{\cal A/G} e^{-S(A)}{\cal D}A \\ 
  \text{converges for $\R$ electromagnetism}
\end{array}\right)
\iff
H^1 = 0
\]

More generally, $p$-form electromagnetism  has what we might call the {\bfm $p$-form Bohm-Aharonov effect}, but whereas the ordinary Bohm-Aharonov effect applies to regions of spacetime with nontrivial {\em first} cohomology, the  $p$-form version depends on the $p$th cohomology.  We have also seen that getting path integrals in $\R$ $p$-form electromagnetism to converge requires killing off all `$p$-form Bohm-Aharonov modes':   
\[
\left(\begin{array}{c}
  Z = \int_{\cal A/G} e^{-S(A)}{\cal D}A \\ 
  \text{converges for $\R$ $p$-form electromagnetism}
\end{array}\right)
\iff
H^p = 0
\]

In fact, when $H^p$ is trivial, we can sometimes simplify the process of taking path integrals further.  To see this, note that the homomorphism $d_p:C^p \to C^{p+1}$ induces an isomorphism
\[
             \frac{C^p}{Z^p} \iso B^{p+1}.
\]
In the case where $H^p$ is trivial, $C^p/Z^p=C^p/B^p$, which is the physical configuration space.  If in addition $d_p$ is {\em onto}, so that $B^{p+1}=C^{p+1}$, we then have an isomorphism 
\[
         \frac{\cal A}{\cal G}=\frac{C^p}{Z^p} \iso C^{p+1}.
\] 
In other words, when there are no Bohm-Aharonov modes, and $d_p$ is surjective, we can just as well do path integrals over the space of curvatures $F$, rather than over the space of $p$-connections.  In our first example above, $H^1$ is trivial and every possible curvature is $d$ of some connection, so we could calculate the partition function more easily as:
\[
    Z= \int_{C^2} e^{-\frac{1}{2e^2}\left({F_1}^2 + {F_2}^2\right)} dF_1 dF_2 = 2\pi e^2
\]
The value of the partition function is different in this case, but this does not matter since $Z$ is just a normalization factor --- what matters is that expected values of observables are the same.

For  $p$-form electromagnetism with gauge group $\R$, we thus have several choices for what space to integrate over when we do path integrals:
\begin{enumerate}
\item $\cal A$, the space of $p$-connections --- This is the na\"ive approach of Section \ref{sec:survey}, and works almost never: only when there are no nontrivial flat connections.
\item ${\cal A}/{\cal G}$, the space of $p$-connections mod gauge transformations --- This works whenever there are no $p$-form Bohm-Aharonov modes, i.e. when spacetime has trivial $p$th cohomology.
\item ${\cal A}/{\cal A}_0$, the space of $p$-connections mod flat $p$-connections.  This always gives convergent path integrals, and is the same as ${\cal A}/{\cal G}$ when the $p$th cohomology is trivial.   However, in the case of nontrivial cohomology, it means ignoring the $p$-form Bohm-Aharonov modes.
\item ${\cal F}:= C^{p+1}$, the space of curvatures of $p$-connections.  In the case where the $p$th cohomology is trivial {\em and} $d\maps C^p \to C^{p+1}$ is onto, this is equivalent to the previous two options, and generally far easier to calculate with.
\item ${\cal F}_0:= B^{p+1}$, the space of curvatures $F$, subject to the constraint $dF=0$.  This is like the previous option, but works even when $d\maps C^p \to C^{p+1}$ is not onto.  The practical tradeoff is that the constraint makes calculating integrals less straightforward.
\end{enumerate}
The major shortcoming in the case of gauge group $\R$ is that none of these options allows us to take any $p$-form Bohm-Aharonov modes into account and get convergent path integrals.  We see in  Section \ref{sec:u1} that this problem is resolved when we switch the gauge group to $U(1)$.  We remark that options 4 and 5, while sometimes convenient in the free field setting, are less viable when we add matter, since matter fields typically couple to the $A$ field, not $F$.

Let us look at some examples.  We have seen that the torus has nontrivial cohomology.  In fact, for purposes of computing cohomology, we could have trimmed down our example to a torus with only one vertex, two edges, and one plaquette:
\[
\xy
(0,0)*{\includegraphics{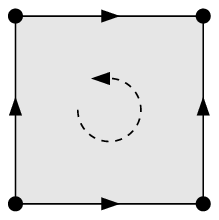}};
(-12,12)*{\displaystyle v_1};
(12,12)*{\displaystyle v_1};
(-12,-12)*{\displaystyle v_1};
(12,-12)*{\displaystyle v_1};
(0,12)*{e_1};
(0,-12)*{e_1};
(-12,0)*{e_2};
(12,0)*{e_2};
(0,0)*{p_1};
\endxy
\]
Since cohomology is a topological invariant --- topologically equivalent spaces have isomorphic cohomology ----we find again the cohomology groups 
\[
     H^0\iso \R \qquad H^1\iso \R^2 \qquad H^2 \iso \R.
\]
In particular, $H^1$ tells us there are two Bohm-Aharonov modes, just as there were for our two-plaquette torus above, and just as there would be for any other spacetime with the topology of a 2-torus.  

As another example, consider a discrete spacetime with the topology of an $n$-sphere, $S^n$.   When the gauge group is $\R$ this has cohomology groups
\[
      H^k(S^n,\R)\iso \left\{
      \begin{array}{cl} 
          \R & k=0 \text{ or } n \\
          0 & \text{otherwise}
      \end{array}\right.
\]
Thus, for example, there are no Bohm-Aharonov modes for ordinary electromagnetism on the 2-sphere, since $H^1(S^2)=0$.  Yet for the sphere the differential $d_1$ is not surjective.  Thus in this case we may do path integrals over the physical configuration space $\cal A/G$, but not over the entire space of curvatures, $C^2$.   By contrast, for a Kalb-Ramond field on the 2-sphere, there are 2-form Bohm-Aharonov modes , since $H^2(S^2)\iso\R$.  In this case, $\R$-electromagnetism gives divergent path integrals unless we kill these modes.   But what about the other nontrivial cohomology, $H^0(S^n)$?  This evidently corresponds to a Bohm-Aharonov effect for `0-form electromagnetism'.   We consider this in Section \ref{sec:zeroform}.

\subsection{Hodge's Theorem, Topological Classical Field Theory}
\label{hodge}

Having quantized lattice $p$-form electromagnetism, let us step back and take a closer look at the classical theory.  
Recall from Section \ref{sec:maxwell} that the vacuum discrete $p$-form Maxwell equations:
\[
  d^\dagger F = 0 \text{ and }
  dF = 0
\]
have two possible interpretations:
\begin{enumerate}
\item Considering the gauge potential $A$ as fundamental, we have seen the positive definiteness of the inner product on $(p+1)$-forms reduces the equations of motion to $F=0$.  From this perspective, then, solutions of the vacuum discrete $p$-form Maxwell equations are fundamentally $p$-forms $A$ satisfying $dA=0$, modulo exact $p$-forms.  In the global language, solutions are precisely elements of the space of flat $p$-connections modulo gauge transformations.  But this is just
\[
  \frac{{\cal A}_0}{\cal G} = \frac{Z^p}{B^p} = H^p,
\]
the $p$th cohomology.

\item Considering the electromagnetic field strength $F$ as fundamental, solutions of the vacuum discrete $p$-form Maxwell equations are $(p+1)$-forms $F$ satisfying
\[
   dF = 0 \text{ and } d^\dagger F= 0.
\] 
The second of these equations says precisely that $F$ is orthogonal to every exact $(p+1)$-form, since the nondegeneracy of the inner product says $d^\dagger F= 0$ if and only if $\langle d^\dagger F, X\rangle= 0$ for every $p$-form $X$, or in other words $\langle F, dX\rangle= 0$ for every exact $(p+1)$-form $dX$.   We show below that positive definiteness of the inner product on $p+1$ forms implies that the subspace of closed $(p+1)$-forms $F$ which are also orthogonal to every exact $(p+1)$-form $dX$, is isomorphic to 
\[
   H^{p+1},
\]
the $(p+1)$st cohomology.
\end{enumerate}
So, from either viewpoint, the classical theory is `topological'; but the two viewpoints disagree on which aspect of the topology determines the solutions!  

To prove the claim in item 2 above, it is helpful to develop an alternate viewpoint on cohomology --- a viewpoint  which is useful whenever we have a cochain complex of (possibly indefinite) inner product spaces.  When we have such a complex $C$, each of the differentials $d_q\maps C^q \to C^{q+1}$ has an adjoint $d_q^\dagger \maps C^{q+1} \to C^q$:
\[
\xymatrix{ C^{q-1} \ar @< 2pt> [r]^{d_{}} & 
                  C^{q} \ar @< 2pt> [l]^{d^\dagger_{}} \ar @< 2pt> [r]^{d} & 
                  C^{q+1} \ar @< 2pt> [l]^{d^\dagger_{}} }
\]
The first important observation is that $dd=0$ together with nondegeneracy of the inner product implies $d^\dagger d^\dagger=0$, so the maps $d^\dagger$ also form a complex.  We say $\omega \in C^q$ is {\bf coclosed} if $d^\dagger\omega=0$ or {\bf coexact} if $\omega= d^\dagger \alpha$ for some $\alpha\in C^{q+1}$.  Just as we noted in the special case of $F$ above, a coclosed $q$-form is precisely a $q$-form which is orthogonal to all exact $q$-forms.  Similarly, a closed $q$-form is precisely a $q$-form which is orthogonal to all coexact $q$-forms, by the same argument.  That is:
\[
   \ker d_q = (\ran d_q^\dagger)^\perp \qquad \ker d_{q-1}^\dagger = (\ran d_{q-1})^\perp.
\]
We shall combine these to prove the {\bf Hodge decomposition}\footnote{There are more sophisticated versions of the Hodge decomposition, such as the usual version on the deRham complex, and the `Kodaira decomposition' for complexes of operators that are only densely defined \cite{Carrion}.  We supply the proof of this simple version here because it is illuminating.  Proofs of fancier versions are the same in essential structure, though some nontrivial analysis is involoved.}  which says we can write $C^q$ as the orthogonal direct sum
\[
    C^q = \ran d_{q-1} \oplus \ker\Delta_q \oplus \ran d_q^\dagger.
\]
Here 
\[
\Delta_q:= d^\dagger_{q} d_{q} + d_{q-1}d^\dagger_{q-1}
\]
 is the {\bf Laplacian} on $C^q$, and a cochain in its kernel is said to be {\bf harmonic}. 

\vspace{1em}

\noindent{\bf Proof of the Hodge Decomposition:}   
The remarks immediately preceding the statement of the Hodge Decomposition give us two separate direct sum decompositions of $C^q$:
\[
 C^q  = \ker d_q \oplus \ran d_q^\dagger =   \ker d_{q-1}^\dagger \oplus \ran d_{q-1}.      
\]
The first of these says we can write a given $\omega\in C^q$ uniquely as $\omega = \omega_0 + d^\dagger \beta$ with $\omega_0 \in \ker d_q$.   We then use the second direct sum to write $\omega_0 = \omega_h + d\alpha$ with $\omega_h \in \ker d_{q-1}^\dagger$.  But in fact, this $\omega_h = \omega_0- d\alpha$ is also in the kernel of $d_q$ since $\omega_0$ and $d\alpha$ both are.  We thus have
\[
    \omega = d\alpha + \omega_h + d^\dagger \beta \qquad 
    \text{with } \omega_h \in (\ker d^\dagger_{q-1})\cap (\ker d_{q})
\] 
Moreover, this decomposition is unique, since  $\omega_h + d^\dagger \beta$ and  $d\alpha$ are uniquely determined by  $C^q = \ker d_{q-1}^\dagger \oplus \ran d_{q-1}$, while $\omega_h + d\alpha$ and $d^\dagger \beta$ are uniquely determined by $C^q= \ker d_q \oplus \ran d_q^\dagger$.  The sum is thus direct:
\[
       C^q = \ran d \oplus (\ker d^\dagger\cap \ker d) \oplus \ran d^\dagger,
\]
and orthogonal since $\omega_h$, $d\alpha$, and $d^\dagger \beta$ are mutually orthogonal.

Now suppose $\omega$ is harmonic.  Then
\begin{align*}
         0 &= \langle \omega,(d^\dagger d + dd^\dagger)\omega\rangle \\
             &= \langle d^\dagger\omega, d^\dagger \omega\rangle + \langle d\omega, d \omega\rangle     . 
\end{align*}
Up to this point, we have not used that the inner product is positive definite.    Using this fact now, we conclude that
\[
d\omega=0 \text{ and } d^\dagger \omega =0.
\]
Conversely, if $\omega$ solves these equations, then obviously $(d^\dagger d + dd^\dagger)\omega=0$, so $\omega$ is harmonic.
We thus see that the space of harmonic $q$-forms is the space of closed and coclosed $q$-forms, and the decomposition is established.  $\blacksquare$ 

Recall that the free $p$-form Maxwell's equations for $F$ simply say that $F$ is closed and coclosed.  The above proof thus shows that when we have a positive definite inner product defined on all $(p+1)$-forms, solutions of Maxwell's equations are precisely the harmonic $(p+1)$-forms.   In fact the Hodge decomposition lets us prove a version of  {\bf Hodge's Theorem}, which says the space of harmonic $q$-forms is canonically isomorphic to the cohomology $H^q$.   To see this, write a given $q$-form $\omega$ as
\[
       \omega = d\alpha + \omega_h + d^\dagger \beta
\]
with $\omega_h$ harmonic.  To understand cohomology in this context, we need to understand what it means for $\omega$ to be closed.   But we have observed that a closed form is precisely a form orthogonal to every coexact form $d^\dagger \mu$.  Since the Hodge decomposition of $\omega$ is orthogonal, this means $\omega$ is closed precisely when its coexact part $d^\dagger \beta$ vanishes. 
Thus, noting that $\ran d_{q-1}$ is just $B^q$, the Hodge decomposition
\[
    C^{q} = B^{q} \oplus \ker \Delta_q \oplus \ran d^\dagger
\]
implies
\[
    Z^{q} = B^{q} \oplus \ker\Delta_q
\]
and hence
\[
   H^{q} :=\frac {Z^{q}}{B^{q}} \iso \ker\Delta_q.
\]
That is, the $q$th cohomology $H^{q}$ is canonically isomorphic to the subspace of harmonic $q$-forms.  Identifying these two spaces, we can therefore write:
\[
      C^{q} = dC^{q-1} \oplus H^{q} \oplus d^\dagger C^{q+1}
\] 
as an alternate version of the Hodge decomposition of $C^q$.

In the case we are most interested in, the case $q=p+1$, we thus see that the solutions of the $p$-form Maxwell equations
\[
      dF = 0 \text{ and } d^\dagger F= 0
\]
are 
\[
  \ker d \cap \ker d^\dagger = \ker\Delta_{p+1}\iso H^{p+1}  .
\]
That is, when we consider $F$ as the fundamental classical field, solutions of the equations of motion are specified by the $(p+1)$st cohomology $H^{p+1}$:
\[
dF = 0 \text{ and } d^\dagger F= 0 \qquad \iff \qquad F\in H^{p+1}.
\]

\section{{\boldmath$U(1)$} {\boldmath $p$}-Form Electromagnetism and the {\boldmath $p$}-form Bohm-Aharonov Effect} 
\label{sec:u1}

{\boldmath
\subsection{Gauge Groups for $p$-form Electromagnetism: $\R$ vs.\ $U(1)$}
}
We have seen that eliminating all of the divergences from path integrals in $\R$ $p$-form electromagnetism requires that we factor out not only all of the gauge freedom, but also the Bohm-Aharonov modes.  This is undesirable: the Bohm-Aharonov effect is an empirical fact in ordinary 1-form electromagnetism.  Neither should we rule out the {\em $p$-form} Bohm-Aharonov effect in our calculations.  The ultimate solution to the problem of divergent path integrals is to switch our gauge group from $\R$ to $U(1)$.  We shall see that $U(1)$ electromagnetism cures all divergences, even in cases where the $p$th cohomology is nontrivial!  Better yet, $U(1)$ path integrals converge even before we factor out the gauge freedom.  The reason for this is simple.  The space of $U(1)$ $p$-connections is 
\[
            {\cal A}(U(1)) = U(1)^{X_p}
\]
where $X_p$ is the set of $p$-cells in the lattice.  But this is just a product of circles --- a torus!  Since a torus is compact, 
\[
   \int_{U(1)^{X_p}} f(A)  e^{-S(A)}{\cal D}A
\]
converges for any continuous function $f$ of $A$.  Of course, for $f$ to be a physical observable, we still want $f$ to be gauge invariant: $f(A) = f(A+d\phi)$.

Indeed, the switch to $U(1)$ promises to be such an improvement that the reader may well wonder why we have bothered with the $\R$ case.   As it turns out, the $U(1)$ theory can be obtained most easily from the $\R$ theory.  In particular, without having developed the $\R$ theory, it would be difficult to guess the best action to use in evaluating $U(1)$ path integrals.  The reason is that since the curvature lies not in the vector space $\R^{X_{p+1}}$ but in the mere group $U(1)^{X_{p+1}}$, we do not have the same analogy between cochains and and differential forms as we had in the $\R$ case.  In particular, we cannot rely directly on an inner product of cochains for the action as we did in the $\R$ case in Section \ref{sec:action}.   What we will show is that we can turn $p$-form electromagnetism with gauge group $\R$ into $p$-form electromagnetism with gauge group $U(1)$, essentially by ``wrapping the real line around the circle."

The key to understanding the relationship between the abelian gauge groups $\R$ and $U(1)$ is the following `short exact sequence' of homomorphisms:
 \[
 \xymatrix{
 0 \ar[r] & {\Z} \ar@{->}[r] & {\R} \ar@{->}[r]& {U(1)} \ar[r] & 0
 }
 \]
where the second map is the usual inclusion and the third sends $x\in\R$ to  $e^{2\pi ix}\in U(1)$.  Here `exactness' means the same as for the chain and cochain complexes: that the range of one map equals the kernel of the next.  In particular, this says we map $\R$ to $U(1)$ by winding the line around the circle once every $2\pi$, and that the kernel of this map is precisely $\Z$.

In comparing different gauge groups, it is convenient to write the group of $k$-form fields not simply as $C^k$ but more explicitly as 
\[
  C^k(G) = \hom (C_k, G),
\]
i.e.\ homomorphisms from the group of $k$-chains to the gauge group $G$.  

Now given any $\R$ $p$-connection
\begin{align*}
           A\maps C_p \to \R \\
\end{align*}
we get a $U(1)$ $p$-connection 
\[
      \hat A:=   e^{2\pi i A}\maps C_p \to U(1)
\]
by composition with the homomorphism $\R\to U(1)$.  We thus get a map
\begin{align*}
  \hom(C_p,\R) &\to  \hom(C_p,U(1)) \\
           {A} &\mapsto {\hat A}
\end{align*}
and the fact that the above sequence of abelian groups is exact implies that the kernel of this map is $\hom(C_p,\Z)$.  Moreover, since $C_p$ is free, every $U(1)$ $p$-connection comes from an $\R$ $p$-connection in precisely this way.  
Given a $p$-connection $A\maps C_p\to U(1)$, map each generator $c$ in the basis $X_p$ of $C_p$ to some $\tilde A(c) \in \R$ with $\exp(2\pi i \tilde A(c)) = A(c)$.  This defines an $\R$ $p$-connection $\tilde A$ which `lifts' $A$:  
\[
  \xymatrix{
   & C_p \ar@{.>}[dl]_{\tilde A} \ar[d]^A \\
\R \ar[r] &U(1) \ar[r]& 0
}.
\]
That is, we have the identity:
\[
    {\hat {\tilde A}} = A.
\]

In fact, 
each of the sequences 
\[
  \xymatrix{
  0\ar[r] & {\hom(C_k,\Z)} \ar[r] &{\hom(C_k,\R)} \ar[r] &{\hom(C_k,U(1))} \ar[r] & 0
  }
\]
is exact, thus giving a short exact sequence of chain maps:
\[
\xymatrix{
0 \ar[r] & {\hom(C_n,\Z)} \ar[r]  & {\hom(C_n,\R)} \ar[r]  & {\hom(C_n,U(1))} \ar[r] & 0 \\
{} & { \vdots} \ar[u]^{d} & { \vdots} \ar[u]^{d} & { \vdots} \ar[u]^{d} \\
0 \ar[r] & {\hom(C_1,\Z)} \ar[r] \ar[u]^d & {\hom(C_1,\R)} \ar[r]\ar[u]^d  & {\hom(C_1,U(1))} \ar[u]^d \ar[r] &0 \\
0 \ar[r] & {\hom(C_0,\Z)} \ar[r] \ar[u]^d & {\hom(C_0,\R)} \ar[r]\ar[u]^d  & {\hom(C_0,U(1))} \ar[u]^d  \ar[r] & 0}
\]  
which we may write more succinctly as
\[
  \xymatrix{
  0\ar[r] & {\hom(C_\bullet,\Z)} \ar[r] &{\hom(C_\bullet,\R)} \ar[r] &{\hom(C_\bullet,U(1))} \ar[r] & 0
  }.
\]
Lattice $p$-form electromagnetism with abelian gauge group $G$ is all about the cochain complex $\hom(C_\bullet, G)$.  We may thus express the physical content of the above exact sequence of chain maps by saying there is a projection from the theory with gauge group $\R$ to the theory with gauge group $U(1)$, as we wanted, and that the kernel of this projection is a theory with gauge group $\Z$.  Metaphorically:
\[
\xymatrix{
0 \ar[r] &   
   {\left( 
   {\xy 
     (0,7.5)*{\text{lattice $p$-form}};
     (0,2.5)*{\text{electromagnetism}};
     (0,-2.5)*{\text{with gauge group}};
     (0,-7.5)*{\Z};
    \endxy} 
    \right)}
   \ar[r] &
   {\left( 
   {\xy 
     (0,7.5)*{\text{lattice $p$-form}};
     (0,2.5)*{\text{electromagnetism}};
     (0,-2.5)*{\text{with gauge group}};
     (0,-7.5)*{\R};
    \endxy} 
    \right)}
   \ar[r] &
   {\left( 
   {\xy 
     (0,7.5)*{\text{lattice $p$-form}};
     (0,2.5)*{\text{electromagnetism}};
     (0,-2.5)*{\text{with gauge group}};
     (0,-7.5)*{U(1)};
    \endxy} 
    \right)}
   \ar[r] &
0  
}  
\]  
is exact.

One critical implication of the exact sequence of cochain complexes is that two $\R$ $p$-connections $A,A'\in \hom(C_p,\R)$ which are gauge equivalent, say 
\[
       A- A' = d\phi \qquad \phi\in\hom(C_{p-1},\R),
\] 
project down to $U(1)$ $p$-connections $\hat A,\hat A'\in\hom(C_p,U(1))$ which are gauge equivalent:
\[
        \hat A- \hat A' = d\hat \phi \qquad \hat\phi\in\hom(C_{p-1},U(1)).
\]
Briefly, $\R$ gauge equivalence implies $U(1)$ gauge equivalence.

But in some cases, there is a genuine difference between the two choices of gauge group.  A simple example is a space which is sometimes called an `$m$-fold dunce cap', $DC_m$.  This can be constructed from one vertex $v$, one edge $e$, and one plaquette $p$ which is sewn along its boundary around the looped edge in such a way that it wraps around $m$ times:
\[
\xy
 (0,0)*{\includegraphics{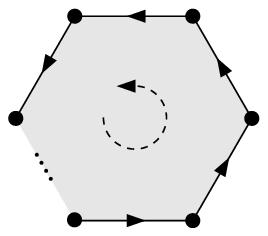}};
 (0,0)*{p};
 (0,-13)*{e};
 (0,12.5)*{e};
 (11,6)*{e};
 (11,-6)*{e};
 (-11,6)*{e};
 (14,0)*{v};
 (-14,0)*{v};
 (-8,11)*{v};
 (8,11)*{v};
 (8,-11)*{v};
 (-8,-11)*{v};
\endxy
\]
The dunce cap has cohomology groups 
\[
    H^0(DC_m, \R) \iso \R \qquad H^1(DC_m, \R) =0 \qquad H^2(DC_m, \R) \iso 0 
\]
for $G=\R$, but 
\[
    H^0(DC_m, U(1)) \iso U(1) \qquad H^1(DC_m, U(1)) \iso \Z/m \qquad H^2(DC_m, U(1)) = 0     
\]
for $G=U(1)$.  The significant difference is in the first cohomology. The fact that $H^1$ is trivial in the $\R$ case says every $\R$-valued 1-cocycle is a 1-coboundary.  Said another way, all flat $\R$ connections on $DC_m$ are gauge equivalent:
\[
  dA=dA' = 0 \implies d(A-A')=0 \implies A-A' = d\phi.
\]
In fact, this result is not surprising given our particular construction of $DC_m$ --- there is only {\em one} flat $\R$ connection, the zero connection.  But in the $U(1)$ case, $H^1$ is a cyclic group of order $m$.  It is not  hard to see why.  There are $m$ flat $U(1)$ connections on $DC_m$, given by assigning to the edge $e$ any one of the $m$ distinct $m$th roots of unity in $U(1)$.  

\subsection{Gaussian Integrals on a Torus}
In the $U(1)$ version of lattice electromagnetism, the holonomy of the connection takes values on the circle $U(1)$.  To do path integrals, we need the analog of a Gaussian on a circle, which we obtain from the ordinary Gaussian by wrapping the real line around the circle.  In particular, since 
\[
\frac{e^{-x^2/2\sigma^2} dx}{\sqrt{2\pi}\sigma}
\]
is a probability measure on $\R$, 
\begin{equation}\label{circlegaussian}
\sum_{n\in \Z} \frac{e^{-(\theta + 2n\pi)^2/2\sigma^2} d\theta}{\sqrt{2\pi}\sigma}
\end{equation}  
is a probability measure on $S^1=U(1)$ 
which we shall refer to as a {\bf circular Gaussian}.  
We can use this measure to calculate the expected value of a function $f\maps U(1)\to \R$ of the random variable $\theta$ as follows:
\begin{align*}
\langle f \rangle &= \int_0^{2\pi} f(\theta)
\sum_{n\in \Z} \frac{e^{-(\theta + 2n\pi)^2/2\sigma^2} d\theta}{\sqrt{2\pi}\sigma} \\
 &= \frac{1}{\sqrt{2\pi}\sigma} \sum_{n\in \Z} 
      \int_{2n\pi}^{2(n+1)\pi} f(\theta) e^{-\theta^2/2\sigma^2} d\theta \\
 &= \frac{1}{\sqrt{2\pi}\sigma}  
      \int_{-\infty}^{\infty} f(\theta) e^{-\theta^2/2\sigma^2} d\theta 
\end{align*}
What this calculation shows is that integrating a function $f\maps U(1)\to \R$ against the measure in (\ref{circlegaussian}) is the same as extending $f$ to a periodic function $f\maps  \R\to \R$ and then integrating this against the usual Gaussian measure. 

Since a periodic function on $\R$ (or equivalently, a function on the circle) can be expanded in a Fourier series, we should work out the expected values for each element of the Fourier basis $\{e^{im\theta}\}$.  For any $m\in \Z$ we have
\begin{align*}
  \langle e^{im\theta}\rangle &= \frac{1}{\sqrt{2\pi}\sigma}  
      \int_{-\infty}^{\infty} e^{im\theta} e^{-\theta^2/2\sigma^2} d\theta \\
      &=  \frac{1}{\sqrt{2\pi}\sigma}  
      \int_{-\infty}^{\infty}  e^{-(\theta^2-2im\sigma^2\theta)/2\sigma^2} d\theta \\
      &=  \frac{1}{\sqrt{2\pi}\sigma}  
      \int_{-\infty}^{\infty}  e^{-(\theta^2-2im\sigma^2\theta-m^2\sigma^4)/2\sigma^2}
                                       e^{-m^2\sigma^2/2} d\theta \\
      &=  \frac{e^{-m^2\sigma^2/2}}{\sqrt{2\pi}\sigma}  
      \int_{-\infty}^{\infty}  e^{-(\theta-im\sigma^2)^2/2\sigma^2}
                                       d(\theta-im\sigma^2) \\
      &=  \frac{e^{-m^2\sigma^2/2}}{\sqrt{2\pi}\sigma} \cdot \sqrt{2\pi}\sigma\\
      &= e^{-m^2\sigma^2/2}.
\end{align*}
Notice that this expected value is real, even though $e^{im\theta}$ is complex.  The imaginary part is zero on average, since the imaginary parts $\sin(m\theta)$ are odd about the circle's origin. 

In fact what we are calling circular Gaussians are actually famous special functions in their own right, though they are usually presented in a slightly different form.  Using the Fourier coefficient $\langle e^{imx}\rangle$ above, one can expand the circular Gaussian itself in the basis $\{e^{im\theta}\}$, resulting in the identity:
\begin{equation}
\label{circ_theta}
\frac{1}{\sqrt{2\pi}\sigma} \sum_{n\in \Z} e^{-(\theta + 2n\pi )^2/2\sigma^2}
 = \frac{1}{2\pi} \sum_{n\in \Z} e^{-n^2\sigma^2/2}e^{in\theta } .
\end{equation}
The form on the right side of this equality makes it easier to recognize that our circular Gaussian with `standard deviation' $\sigma^2$ is really:
\[
  \frac{1}{2\pi} \vartheta\left(\frac{\theta }{2\pi},i\frac{\sigma^2}{2\pi}\right)
\]
where
\[
\vartheta(z,\tau)= \sum_{n\in\Z}e^{\pi i n^2 \tau}e^{2\pi inz}
\]
is the {\bf theta function} \cite{Mumford}.

When it is convenient, we continue to refer to  this theta function loosely as a `circular Gaussian', and write it as ``$e^{-\theta^2/2\sigma^2}$" even though this is not technically the exponential of anything.  We engage in this gross abuse of notation because it is suggestive of how one actually  calculates {\em integrals} using the corresponding `theta measure' --- by expanding the circle to a line and integrating an ordinary Gaussian.

The higher dimensional version of the above calculations follow the same form.  Suppose we have a Gaussian
\[
     e^{-\half x\cdot Q x  } = e^{- \half Q^{ij}x_i x_j }
\]
on $\R^N$.  Following the prescription we used for the circular Gaussian above, we wrap each of the $N$ coordinates around a circle:
\[
      \sum_{n\in \Z^N}e^{-\half Q^{ij}(\theta_i+2\pi n_i)(\theta_j+2\pi n_j)}      
\]  
to obtain an analog of a Gaussian on a torus, what we might call a {\bf toroidal Gaussian}.
Then:
\begin{align*}
  \langle e^{im^k\theta_k}\rangle &= \sqrt{\frac{\det(Q^{ij})}{(2\pi)^N}}  
      \sum_{n\in \Z^N}\int_{U(1)^N} e^{im^k\theta_k}  
      e^{-\half Q^{ij}(\theta_i+2\pi n_i)(\theta_j+2\pi n_j)} d\theta\\      
      &= \sqrt{\frac{\det(Q^{ij})}{(2\pi)^N}}
      \int_{\R^N} e^{im^k\theta_k} e^{-\half Q^{ij}\theta_i\theta_j} d\theta \\
      &=  \sqrt{\frac{\det(Q^{ij})}{(2\pi)^N}}
      \int_{\R^N}  e^{-\half Q^{ij}(\theta_i\theta_j -2iQ^{-1}_{jk} m^k\theta_i )} d\theta \\
      &= \sqrt{\frac{\det(Q^{ij})}{(2\pi)^N}} e^{-\half Q^{-1}_{kl}m^km^l}   
      \int_{\R^N}  e^{-\half Q^{ij}(\theta_i\theta_j -2iQ^{-1}_{jk} m^k\theta_i -Q^{-1}_{ik}Q^{-1}_{jl}m^km^l)}
                                       d\theta \\
      &= \sqrt{\frac{\det(Q^{ij})}{(2\pi)^N}} e^{-\half Q^{-1}_{kl}m^km^l}   
      \int_{\R^N}  e^{-\half Q^{ij}(\theta_i -iQ^{-1}_{ik} m^k)(\theta_j -iQ^{-1}_{jl} m^l)}
                                       d\theta \\
      &= e^{-\half Q^{-1}_{kl}m^km^l}  .
\end{align*}

Using this and proceeding as in the 1-dimensional case, we obtain the identity
\begin{equation}
\label{tor_theta}
   \sqrt{\frac{\det Q}{(2\pi)^N}}
    \sum_{n\in \Z^N}e^{-\half Q^{ij}(\theta_i+2\pi n_i)(\theta_j+2\pi n_j)} 
    = \frac{1}{(2\pi)^N} 
    \sum_{n\in \Z^N} e^{-\half Q^{-1}_{ij}n^i n^j} e^{in^k\theta_k},
\end{equation}
which reduces to (\ref{circ_theta}) in the case $N=1$, $Q=1/\sigma^2$.  
The right side is again a {\bf theta function}, this time its multivariable generalization:
\[
        \frac{1}{(2\pi)^N} \vartheta \left(\frac{\theta}{2\pi}, i\frac{Q^{-1}}{2\pi} \right)
\]
 Here
\[
            \vartheta( z, \Omega):= \sum_{n\in \Z^N} e^{\pi i n\cdot \Omega n + 2\pi i n\cdot z}
\]
where $z\in \C^N$, and $\Omega$ is a symmetric $N\times N$ complex matrix with positive definite complex part.\cite{Mumford}  Again, we continue to write this theta function as ``$e^{-\half\theta \cdot Q\theta}$" and think of it as a `toroidal Gaussian' when convenient, since this is evocative of the analogy to Gaussians on $\R^N$.

\subsection{$U(1)$ Path Integrals}
\label{u1path}

Recall that for gauge group $\R$, our action for discrete $p$-form electromagnetism is given by
\[
       S(A) = \frac{1}{2e^2} \langle F, F \rangle = \frac{1}{2e^2} h^{ij} F_i F_j
\]
where $h^{ij}$ is the matrix of the inner product relative to the basis of $(p+1)$-cochains consisting of dual $(p+1)$-cells.  As discussed in the previous subsection, making the transition from $\R$ to $U(1)$ involves replacing the ordinary Gaussian $\exp(-S)$ in the real variables $F_i$ by a `toroidal Gaussian' 
in the $U(1)$-valued variables $F_i$: 
\begin{equation}
\label{u1S}
\begin{split}
  e^{-S(A)} &= 
                          \sum_{n\in \Z^{N}} e^{-\frac{1}{2e^2} h^{ij} (F_i - 2n_i \pi)(F_j - 2n_j \pi)} \\
                   &= 
                          \sum_{n\in \Z^{N}} e^{-\frac{1}{2e^2} h^{ij} (A(\partial x_i) - 2n_i \pi)(A(\partial x_j) - 2n_j \pi)}
\end{split}
\end{equation}
where $N=|X_{p+1}|$ is the number of $(p+1)$-cells in the spacetime, which we now assume to be finite.   Note that we do not define the {\em action} $S$ but only the analog of its exponential $e^{-S}$,  which is all we need for doing path integrals.

For the sake of comparison, it is perhaps best at this point to return to our old two-plaquette spacetime from Section \ref{sec:survey}:
\[
\usebox{\twoface}
\]
The corresponding chain complex is still as constructed in Section \ref{ncomplex} (and later used in Section \ref{cohomology1}), but our gauge field $A$ is now a $U(1)$-valued 1-cochain:
\[
      A \in \hom(C_1,U(1)).
\]
Thinking of $U(1)$ as the real numbers mod $2\pi$,   if 
\[
    [A_i]\in \R/2\pi\Z \iso U(1)
\]
is the value of $A$ on the edge $e_i$, then using the action from Section \ref{pplus1} in equation (\ref{u1S}) we obtain 
\[
  e^{-S(A)} = \sum_{n\in \Z^2} e^{-(A_2-A_1-2n_1\pi)^2/2e^2V_1 -(A_3-A_2-2n_2\pi)^2/2e^2V_2}
\]
where $A_i$ is any representative of the class $[A_i]$.  Conveniently, the fact that the matrix of the inner product is diagonal in this case lets us split $e^{-S}$ up into one term for each plaquette:
\[
    e^{-S(A)} = e^{-S(A|_{p_1})}e^{-S(A|_{p_2})}
\]
where
\begin{align*}
   e^{-S(A|_{p_1})} &= \sum_{n\in \Z} e^{-(A_2-A_1-2n\pi)^2/2e^2V_1} \\
   e^{-S(A|_{p_2})} &= \sum_{n\in \Z} e^{-(A_3-A_2-2n\pi)^2/2e^2V_2}.
\end{align*}
Note the correspondence: since the path integral in the gauge group $\R$ case involves a product of Gaussians, one for the holonomy around the boundary of each plaquette, the path integral for the $U(1)$ case involves a product of theta functions --- the `circular Gaussians' we defined above.  

We are ready to calculate the partition function for our example:
\begin{align*}
  Z &= \int_0^{2\pi}\int_0^{2\pi}\int_0^{2\pi}
                  \sum_{n\in \Z} e^{-(A_2-A_1-2n\pi)^2/2e^2V_1}
                  \sum_{m\in \Z} e^{-(A_3-A_2-2m\pi)^2/2e^2V_2} dA_3 dA_2 dA_1 \\
     &= \int_0^{2\pi}\int_0^{2\pi} 
                  \sum_{n\in \Z} e^{-(A_2-A_1-2n\pi)^2/2e^2V_1}
                  \int_{-\infty}^{\infty} e^{-(A_3-A_2)^2/2e^2V_2} dA_3 dA_2 dA_1 \\     
     &= \sqrt{2\pi V_2}e \int_0^{2\pi}\int_0^{2\pi} 
                  \sum_{n\in \Z} e^{-(A_2-A_1-2n\pi)^2/2e^2V_1}
                   dA_2 dA_1 \\ 
     &= \sqrt{2\pi V_2}e \int_0^{2\pi} 
                   \int_{-\infty}^{\infty} e^{-(A_2-A_1)^2/2e^2V_1}
                   dA_2 dA_1 \\ 
      &=(2\pi e)^2\sqrt{V_1 V_2}
\end{align*}
The partition function is finite, in contrast to the $\R$ case.  

In the $U(1)$ case there is actually a nice geometric interpretation of the coupling constant $1/e^2$.  Looking back carefully at the relationship between Gaussians and theta functions, we see that the square of the electric charge may be reinterpreted as the {\em radius} of the circle $U(1)$.  If we take this radius to be effectively infinite, so that the circle becomes a line, then we are back in the realm of $\R$ electromagnetism with {\em unit} electric charge.  This helps explain the factor $(2\pi e)^2$ in our result for the partition function above: as the electric charge becomes infinite,  the partition function effectively diverges, as we verified is Section \ref{sec:survey} that it does in the $\R$ case.  

It is worth noting that the split of $e^{-S}$ into the product $\prod e^{-S(p_i)}$ of plaquette actions happens more generally whenever the inner product is diagonal, and in particular in the case of $p$-form electromagnetism in $p+1$ dimensions.   The partition function for $U(1)$ $p$-form lattice electromagnetism in the special case of $p+1$ dimensions is thus  
\begin{align*}
    Z &= \int_{U(1)^{X_p}} \prod_{c\in C^{p+1}}  e^{-S(A|c)} {\cal D}A \\
       &=\int_{U(1)^{X_p}}  \prod_{c\in C^{p+1}} \sum_{n \in\Z} e^{-(A(\partial c)-2n\pi )^2/2e^2\vol(c) } {\cal D}A.
\end{align*}

\section{The Free Scalar Field and 0-Form \mbox{Electromagnetism}}
\label{sec:zeroform}
The alert reader may have noticed that the formalism for lattice $p$-form electromagnetism developed in the previous sections does not seem to apply equally to the case $p=0$.  In particular, there seems to be no notion of gauge invariance in the 0-form case.  By extending our notion of $n$-complex we now correct this discrepancy.  Lest the reader think this an esoteric digression of purely mathematical interest, we note that $0$-form electromagnetism is really just scalar field theory!  Indeed, the $0$-form Maxwell equation
\[
        d^\dagger d \phi = 0
\] 
really just says, in the Riemannian signature we have been using, 
\[
       \nabla^2 \phi = 0,
\]
or in the more familiar Lorentzian case:
\[
      \Box \phi = 0.
\]

The field $\phi$ in free scalar field theory --- the `gauge field', though the theory is too simple to be considered a `gauge theory' by most --- is just a $G$-valued 0-cochain in the lattice context:
\[
     \phi \in \hom(C_0,G),
\] 
i.e.\ it assigns to each vertex an element of the gauge group $G=\R$ or $U(1)$.  But this is just what we called a `gauge transformation' in electromagnetism.   To figure out $0$-form electromagnetism, let us therefore think again of our motivating example --- ordinary $1$-form electromagnetism, taking the gauge group to be $\R$ again for simplicity.

One key point about gauge transformations in electromagnetism is that different connections can induce the same curvature.  But further thought along these lines reveals that this phenomenon is not at all unique to connections.  Indeed, if we move one notch down the chain, different scalar potentials may give rise to the same connection.  Texts on electromagnetic theory usually summarize this observation by saying the potential is only defined up to an additive constant.  In fact, adding a constant to the potential is like ``gauge transforming the potential."  Better yet, it is like making a ``meta-gauge transformation between gauge transformations."  People don't usually talk about a choice of origin of the potential in these terms, perhaps because it is too simple, but this really is a primitive example of the `ghosts of ghosts' one might read about in string theory papers. 

To make this idea precise, we use the relationship between the connection $A$ and potential (or scalar field) $\phi$ as a model for that between potential and what we might call {\bf pre\-potential}.  This seems to demand we add a new term, some $C^{-1}$, to our cochain complex, along with a new differential $d_{-1}\maps C^{-1}\rightarrow C^{0}$.  Then a meta-gauge transformation of a scalar potential $\phi$ by a meta-potential $u\in C^{-1}$ should look like:
\[
\phi\mapsto \phi + d_{-1}u.
\]

What should $C^{-1}$ and $d_{-1}$ look like?  Usually we only think of adding a constant to the potential, but that is really only because of our peculiar fondness of connected spacetimes.  In general, we can add a {\em local} constant to $\phi$ without affecting $d\phi$, so $C^{-1}$ can have as many degrees of freedom as there are connected components in the lattice.  In fact, we might as well take $C_{-1}$ to be the free abelian group generated by the set of connected components of the lattice, and let $C^{-1}=\hom(C_{-1},\R)$ be its dual.  

Now to make $\phi\mapsto \phi + d_{-1}u$ act as an adjustment of the origin of $\phi$ in each connected component, we simply define $d_{-1}u\maps C_0\rightarrow \R$ to be the map which assigns to each vertex (i.e.\ each basis element) in $C_0$ the value of $u\maps C_{-1}\rightarrow
\R$ on the connected component in which that vertex lies.  That is, given a vertex $v \in C_0$, if we denote its connected component by $\partial v$, then $d_{-1}u(v):=u(\partial v)$.  This, of course, defines a new boundary map $\partial_{-1}\maps C_0\rightarrow C_{-1}$ --- it is just the adjoint of $d_{-1}$.

Putting all of this together, we arrive at a new {`augmented'} chain complex and cochain complex
\begin{align*}
0 \buildrel \partial_{-1} \over \longleftarrow
&C_{-1} \buildrel \partial_{0} \over \longleftarrow
C_{0} \buildrel \partial_{1} \over \longleftarrow
C_{1} \buildrel \partial_{2} \over \longleftarrow
\cdots \buildrel \partial_{n} \over \longleftarrow
C_{n} \\
&C^{-1} \buildrel d_{-1} \over \longrightarrow
C^{0} \buildrel d_{0} \over \longrightarrow
C^{1} \buildrel d_{1} \over \longrightarrow
\cdots \buildrel d_{n-1} \over \longrightarrow
C^{n} \buildrel d_{n} \over \longrightarrow
0
\end{align*}

What we have just seen is that the natural meaning of the notion `boundary of a vertex' is just the connected component in which the vertex lies.   It is interesting to note that at this level the old motto ``the boundary of a boundary is zero" just says that the two endpoints of an edge lie in the same connected component!  As a final remark, note that we cannot continue extending the complex further left.  Since each connected component contains (i.e.\ is the boundary of) some vertex, $\partial_0$ is onto, so $\partial\partial=0$ implies the range of $\partial_{-1}$ must be trivial.  Phrased from the dual perspective, there can be no nontrivial gauge transformations in `$(-1)$-form electromagnetism'.

\section{Discretization Independence in {\boldmath $p+1$} Dimensions}
\label{sec:vft}
Recall from Section \ref{pplus1} that when the gauge group is $\R$ and spacetime is $(p+1)$-dimensional, the action in lattice $p$-form electromagnetism in $p+1$ dimensions takes a particularly simple form.  Namely, it just involves the sum of $F^2= (dA)^2$ over all $(p+1)$-cells, weighted by the volume of each cell. 
\[
       S(A) = \frac{1}{2e^2} \sum_{x\in X_{p+1}} \frac{F(x)^2}{{\rm vol}(x)}
\]
In this case we have the beautiful result that the theory does not depend on the details of the discretization!  In fact, there is a strong sense in which the quantum theory of $p$-form electromagnetism in $p+1$ dimensions is almost topological:  it's only nontopological degree of freedom is the {\em total volume} of spacetime.  It is thus what we might call a {\bf volumetric quantum field theory}. 

To begin seeing why the $p$-form theory in $(p+1)$ dimensions is volumetric,  consider splitting a $(p+1)$-cell of volume $V$ in two by slicing through it with a new $p$-cell, leaving some fraction of the original volume on either side.  Schematically:
\[
\xy
(-20,0) *{
	\xy
	(0,0)*\xycircle(10,10){-};
	(0,0)*{V};
	\endxy};
{\ar (-5,0)*{}; (5,0)*{}};
(20,0) *{
	\xy
	(0,0)*\xycircle(10,10){-};
	(-10,0)*{\bullet};(10,0)*{\bullet} **\dir{-};
	(0,5)*{V_1};
	(0,-5)*{V_2};
	(0,-13)*{V_1+V_2 = V};
	(0,13)*{\phantom{V_1+V_2 = V}};
	\endxy}
\endxy
\] 
We wish to compare the curvature $F_0$ on the original $(p+1)$-cell to the sum $F= F_1 + F_2$ in the finer discretization.  According to the action above, the curvature on the original cell is a Gaussian random variable with probability measure
\[
 \mu_{F_0}=\frac{e^{-F_0^2/V}}{\sqrt{\pi V}} dF_0
\]
where we suppress the coupling constant $1/2e^2$ in this section to avoid cluttering our calculations unnecessarily.

On the other hand, the joint probability measure for the random variables $F_1$ and $F_2$ is
\[
   \mu_{F_1,F_2} =\frac{e^{-F_1^2/V_1}e^{-F_2^2/V_2}}{\pi\sqrt{V_1 V_2}} dF_1 dF_2.
\] 
Making the change of variables
\[
\begin{array}{ll}
F=F_1+F_2\\
G=F_1-F_2
\end{array}
\]
we can rewrite this as a measure for $F$ and $G$:
\[
   \mu_{F,G} =\frac{ e^{-\frac{1}{4}(F+G)^2/V_1}e^{-\frac{1}{4}(F-G)^2/V_2}}
                                  {2 \pi\sqrt{  V_1 V_2}} dF dG.
\] 
where the $2$ in the denominator comes from the Jacobian 
\[
 \left| \frac{\partial(F_1,F_2)}{\partial(F,G)} \right|= \frac{1}{2}.
\]

To get a measure for $F$, we now have only to integrate out the $G$-dependence from $\mu_{F,G}$.  Collecting terms, using $V=V_1+V_2$, and then completing the square in $G$, we get
\begin{align*}
   \mu_{F,G} 
                      &=\frac{e^{-F^2/V}}{2 \pi\sqrt{  V_1 V_2}}
                            e^{-\fourth \frac{V}{V_1V_2}(G-\frac{V_2-V_1}{V}F)^2} dF dG
\end{align*}
Integrating out the dependence on $G$ then gives a normalized measure for $F = F_1+F_2$:
\[
\mu_F 
=\frac{e^{-F^2/V}}{\sqrt{\pi V}}dF,
\]
which is the same as $\mu_{F_0}$.  

What this calculation shows is that we are free to rediscretize our spacetime by knocking down or inserting new $(p+1)$-cells, and that we will always get the same results, as long as we only ask questions that can be asked for either discretization.  In the end, the only degree of freedom besides the topology should be the total volume of the spacetime.  This result could be made more precise.  To complete the argument, one would need to establish a complete set of `moves' which allow passage from one discretization to any other, and then show that the theory is invariant under these moves, as we have done for the move described above.  Since it is special to the case of $p+1$ dimensions, however, we instead focus on a more general description which works in arbitrary dimensions.  This is the subject of the next section. 

\section{Chain Field Theory}
\label{sec:chain}
Electromagnetism is not a topological quantum field theory.  In Section \ref{sec:vft} we began to see that the special case of pure $p$-form electromagnetism in $(p+1)$ dimensions is {\em almost} topological --- it requires only one nontopological datum: the volume of spacetime.   But it is interesting to see to what extent our more general theory of $p$-form electromagnetism in $n$-dimensional spacetime (including ordinary electromagnetism in 4 dimensions) can be given a TQFT-like description.  If we pick an inner product on $\R$-valued $(p+1)$-cochains and define the action as in Section \ref{sec:action}, then this goal can indeed be realized to a large extent.  Before diving in, we warn the reader that this section requires a slight jump in mathematical sophistication.  In particular, we need to assume an acquaintance with category theory, referring the reader to reference \cite{MacLane} for details not presented here.

In topological quantum field theory, `space' at any given time is a compact $(n-1)$-dimensional manifold, whereas spacetime is an $n$-dimensional cobordism connecting two slices of space.  This is easiest to visualize when $n=2$, where space is necessarily just a union of circles, 
 and a cobordism between one slice of space and another 
is a 2-manifold with boundary\footnote{More properly, a cobordism is a diffeomorphism class of such 2-manifolds with boundary.} having these two slices of space as the two components of its boundary,  like this:
\newsavebox{\cob}
\savebox{\cob}{ 
  \includegraphics[height=1.6in]{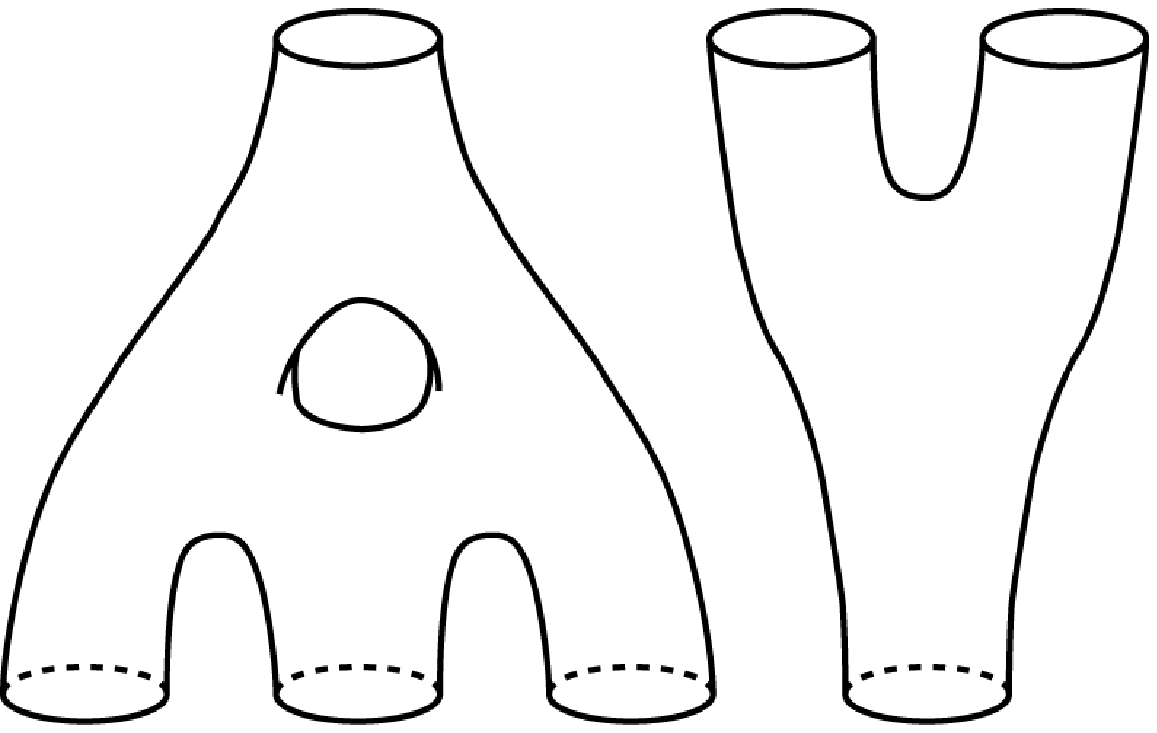}
}
\[
\xy
(0,0)*{\usebox{\cob}};
(40,20)*{S};
(40,-20)*{S'};
(40,0)*{M};
\endxy
\]
A TQFT assigns to each $(n-1)$-manifold $S$ representing space, its Hilbert space of states $Z(S)$, and to each cobordism $M\maps S\to S'$ between $(n-1)$-manifolds a linear operator $Z(M)\maps Z(S)\to Z(S')$, which we can think of as a `time evolution' operator.

More precisely, a TQFT is a symmetric monoidal functor \cite{MacLane}
\[
        Z\maps  \nCob \to \Hilb
\]
from the category of smooth $n$-dimensional cobordisms to the category of Hilbert Spaces.

Since we've described discrete spacetime in electromagnetism using chain complexes, we would like to  define a {\bf chain field theory} to be a symmetric monoidal functor
\[
      Z\maps \nChain \to \Hilb
\]
where $\nChain$ is a category patterned after the category $\nCob$, but with manifolds replaced by chain complexes.  However, as we have already noted, electromagnetism is not merely topological.  Consequently, the category $\nChain$ will have a richer structure than $\nCob$.  In particular, 
$p$-form electromagnetism as described in the previous sections relies on having an inner product on $(p+1)$-forms.

Accordingly, an object $S$ in the category  $\nChain$ should be an $(n-1)$-complex (in the sense of Definition \ref{def:ncomplex}):\footnote{Here we begin a notational practice which we continue throughout this section.  Since we will be dealing with multiple complexes, representing both space and spacetime, we denote, for example, $C_k(S,\R)$ or $C_k(S,U(1))$ by the shorthand $S_k$ rather than $C_k$.}
\[
\xymatrix{
0 & S_0 \ar[l] & S_1 \ar[l] & {\cdots} \ar[l] & S_{n-1} \ar[l] & {\phantom{S_n}}
}
\]
equipped with an inner product on $\R$-valued $(p+1)$-cochains:
\[
     \langle \phantom{\psi},\phantom{\psi} \rangle_{S} \maps  \hom(S_{p+1},\R)\times\hom(S_{p+1},\R) \to \R.
\]
To make our path integrals well defined, we demand that an object $S$ have finitely many cells, so that, in particular, the dimension of $S_{p+1}$ is finite.

To define a morphism from $S$ to $S' \in \nChain$, let $M$ be an $n$-complex, also with finitely many cells, whose chain complex $M_\bullet$ is equipped with injective chain maps 
\[
\xymatrix{
S_\bullet \ar[r]^i & M_\bullet & S'_\bullet  \ar[l]_{i'}
}
\]
sending basis elements to basis elements.  
That is, for each $k\in \{0,1,\ldots,n-1\} $ we have subgroup inclusions $i\maps S_k \to M_k$ and $i'\maps S'_k\to M_k$ such that the following diagram is commutative.
\[
\xymatrix{
0 & S_0 \ar[l] \ar[d]_i & S_1 \ar[l] \ar[d]_i & {\cdots} \ar[l] & S_{n-1} \ar[l] \ar[d]_i  \\
0 & M_0 \ar[l] & M_1 \ar[l] & {\cdots} \ar[l] & M_{n-1} \ar[l] & M_n \ar[l] \\
0 & S'_0 \ar[u]^{i'} \ar[l] & S'_1 \ar[u]^{i'} \ar[l] & {\cdots} \ar[l] & S'_{n-1} \ar[u]^{i'} \ar[l]
}
\]
We also equip $M$ with an inner product on $\R$-valued $(p+1)$-cochains:
\[
     \langle \phantom{\psi},\phantom{\psi} \rangle_{M} \maps  \hom(M_{p+1},\R)\times\hom(M_{p+1},\R) \to \R.
\]
and demand that $M$ preserve the inner products from $S$ and $S'$ in the following sense.  
First, since $M_{p+1}$ is free, we can think of $S^{p+1}= \hom(S_{p+1},\R)$ as a subgroup of $M^{p+1}=\hom(M_{p+1},\R)$ simply by extending  any $\omega \in S^{p+1}$ trivially to $M^{p+1}$.  That is, $\omega$ becomes a member of $M^{p+1}$ by letting it be zero on $M^{p+1}\backslash S^{p+1}$.   We then demand that the inner product $\langle -,- \rangle_{M}$ is precisely $\langle -,- \rangle_{S}$ when restricted to $S^{p+1}\times S^{p+1}$.

A morphism in $\nChain$ will be defined as a certain equivalence class of the gadgets described in the previous paragraph.  Suppose we have two such gadgets from $S$ to $S'$: 
\[
\xymatrix{
S_\bullet \ar[r]^i & M_\bullet & S'_\bullet  \ar[l]_{i'}
}
\]
and
\[
\xymatrix{
S_\bullet \ar[r]^j & M'_\bullet & S'_\bullet  \ar[l]_{j'}
}
\]
We say $M$ and $M'$ are equivalent, written $M\sim M'$, if there exists an isomorphism of chain complexes 
\[
      \phi \maps  M_\bullet \to M'_\bullet 
\] 
preserving both the chosen bases of the $M_k$ and the inner product on $M^p$, such that the diagram
\[
\xymatrix{
S_\bullet \ar[r]^i \ar[dr]_j & M_\bullet \ar[d]_\phi & S'_\bullet \ar[l]_{i'} \ar[dl]^{j'} \\
&                               M'_\bullet
}
\]
 commutes.  
Finally we define a morphism 
 \[
        [M]\maps S \to S'
 \]
 in $\nChain$, which we call a {\bf chain cobordism}, to be the equivalence class under the relation $\sim$ of all gadgets $S_\bullet \longrightarrow M_\bullet \longleftarrow S'_\bullet$.  Following standard practice, we will generally drop the brackets from $[M]$, writing simply $M$ for our chain cobordism, blurring the distinction between the equivalence class and a representative.

Now if $M\maps S \to S'$ and $N\maps  S' \to S''$ are chain cobordisms, we must describe a rule for composing them.   Our mental picture of composing two cobordisms between manifolds as stacking them on top of each other
suggests composing $M$ and $N$ by stacking them to form the larger diagram
\[
\xymatrix{
0 & S_0 \ar[l] \ar[d]_i & S_1 \ar[l] \ar[d]_i & {\cdots} \ar[l] & S_{n-1} \ar[l] \ar[d]_i  \\
0 & M_0 \ar[l] & M_1 \ar[l] & {\cdots} \ar[l] & M_{n-1} \ar[l] & M_n \ar[l] \\
0 & S'_0 \ar[u]^{i'} \ar[l] \ar[d]_j & S'_1 \ar[u]^{i'} \ar[l] \ar[d]_j & {\cdots} \ar[l] & S'_{n-1} \ar[u]^{i'} \ar[l] \ar[d]_j \\
0 & N_0 \ar[l] & N_1 \ar[l] & {\cdots} \ar[l] & N_{n-1} \ar[l] & N_n \ar[l] \\
0 & S''_0 \ar[u]^{j'} \ar[l] & S''_1 \ar[u]^{j'} \ar[l] & {\cdots} \ar[l] & S''_{n-1} \ar[u]^{j'} \ar[l]
}
\]
Of course, while this image is helpful conceptually, this diagram does not give us a new chain cobordism according to the definition given above.  When we compose two cobordisms between manifolds, we don't {\em just} stack them on top of each other; we must glue their boundaries together in such a way that we get a new cobordism --- in particular, a manifold with boundary!  Similarly, when we compose our chain cobordisms  $M\maps S\to S'$ and $N\maps S'\to S''$, we must get a new cobordism $NM\maps S\to S''$
\[
\xymatrix{
S_\bullet \ar[r]^i & (NM)_\bullet & S''_\bullet  \ar[l]_{i'}  
}
\]
by ``gluing $M$ and $N$ together along $S$."  Thinking of $M$ and $N$ as the physical lattices whence they derive,  we want $NM$ to have as $n$-cells all of the $n$-cells of both $M$ and $N$.   
This leads us to define
\[
(NM)_n := M_n \oplus N_n. 
\]
Likewise, $NM$ should have all lower dimensional faces of both $M$ and $N$, except that we identify cells in $M$ and $N$ if they come from the same cell in $S$ under the gluing maps
\[
\xymatrix{
 M_\bullet & S'_\bullet  \ar[l]_{i'}
 \ar[r]^j & N_\bullet 
}.
\]
The natural way to accomplish this is to let 
\[
   (NM)_k := \frac{M_k \oplus N_k}{\langle i'(c) - j(c) | c\in S'_k\rangle} \qquad k= 0, 1, \ldots, n-1.
\]
In words, each $NM_k$ is the direct sum $M_k\oplus N_k$ {\em modulo} relations that say $c\in M_k\subset M_k\oplus N_k$ and $c'\in N_k \subset M_k\oplus N_k$ are equivalent if they are equal as elements of $S'_k$.   In particular, the diagram
\begin{equation}
\label{pushout}
\xymatrix{
S'_\bullet \ar[r]^j \ar[d]_{i'} & N_\bullet \ar[d] \\
M_\bullet \ar[r] & (NM)_\bullet }
\end{equation}
commutes, where the unlabeled arrows are the obvious injections.  
Moreover this is the {\em best} way to make such a commutative square, in the sense that if $K_\bullet$ is any other chain complex of abelian groups equipped with chain maps $\alpha$ and $\beta$ such that 
\[
\xymatrix{
S'_\bullet \ar[r]^j \ar[d]_{i'} & N_\bullet \ar[d]^\alpha \\
M_\bullet \ar[r]_\beta & K_\bullet }
\]
commutes, then there exists a unique chain map $(NM)_\bullet \to K_\bullet$ making 
\[
\xymatrix{
S'_\bullet \ar[r]^j \ar[d]_{i'} & N_\bullet \ar[d] \ar@/^/[ddr]^\alpha \\
M_\bullet \ar[r] \ar@/_/[drr]_\beta & (NM)_\bullet \ar@{.>}[dr] \\
 & & K_\bullet}
\]
commute.  Category theorists express this `universal property' of $NM$ by saying the diagram (\ref{pushout}) is a {\bf pushout square} \cite{MacLane}.  We mention this here because the pushout property streamlines the proofs of Theorems \ref{nChain} and \ref{chainFT}, below.
Note that the $(NM)_k$ are still {\em free}, since the direct product of free abelian groups is free and identifying two basis elements of a free abelian group gives a free abelian group.  

We must define the inner product on $(p+1)$-forms on the composite $NM$.  Since $NM$ consists of all of the cells of $N$ and $M$, there is a canonical way to do this.  Given $F,G \in \hom((NM)_{p+1},\R)$, we define:
\begin{equation}
\label{comp-ip}
    \langle F,G\rangle_{NM} = \langle F|_N, G|_N \rangle_{N} + \langle F|_M, G|_M \rangle_{M}
                                                    - \langle F|_{S'}, G|_{S'} \rangle_{S'}
\end{equation}
Here the first two terms are just the inner products on $M$ and $N$, while the third term corrects for double counting on the common boundary $S'$ of $M$ and $N$.  Specifically, if $x_i$ and $x_j$ are both $(p+1)$-cells in $S'$, then since $S'$ is a subcomplex of both $M$ and $N$, each of the first terms in the inner product contains a term of the form $\langle x^i,x^j\rangle$.  Subtracting $\langle F|_{S'}, G|_{S'} \rangle_{S'}$ thus eliminates the extra $\langle x^i,x^j\rangle$ term.

The reader already convinced, or else willing to accept on faith, that $\nChain$ as we have described it does indeed form a category may skip ahead to Theorem \ref{chainFT}, where we prove the main result.  For the more rigorously oriented, we collect some of the remaining technical details in the proof of the following theorem.

\begin{thm} \label{nChain}
$\nChain$ is a symmetric monoidal category.
\end{thm}

\noindent{\bf Proof:}   To show $\nChain$ is a category, we must show that composition of morphisms is well-defined (i.e.\ that it respects equivalence classes), that it is associative, and that identity morphisms exist.   
  
Suppose $M\sim M'$ are two representatives of a chain cobordism from $S$ to $S'$, and $N\sim N'$ representatives of a cobordism from $S'$ to $S''$.  We must show that $NM\sim N'M'$.  By the equivalence relation $\sim$, there exist basis-preserving isomorphisms  
\[
      \phi \maps  M_\bullet \to M'_\bullet  \qquad  \psi\maps  N_\bullet \to N'_\bullet
\] 
such that the following diagrams commute:
\[
\xymatrix{
S_\bullet \ar[r]^i \ar[dr]_j & M_\bullet \ar[d]_\phi & S'_\bullet \ar[l]_{i'} \ar[dl]^{j'} \\
&                               M'_\bullet
}
\qquad 
\xymatrix{
S'_\bullet \ar[r]^k \ar[dr]_\ell & N_\bullet \ar[d]_\psi & S''_\bullet \ar[l]_{k'} \ar[dl]^{\ell'} \\
&                               N'_\bullet
}
\]
Here is where pushouts come in handy.  By the fact that $NM$ and $N'M'$ are both pushouts, there exist unique morphisms $\sigma$ and $\tau$ making the following diagrams commute:
\[
\xymatrix{
S'_\bullet \ar[r]^k \ar[d]_{i'} \ar@/^2em/[rr]^\ell \ar@/_2em/[dd]_{j'}
         & N_\bullet \ar[r]^\psi \ar[d] & N'_\bullet \ar[dd] \\
M_\bullet \ar[r]\ar[d]_\phi & (NM)_\bullet \ar@{.>}[dr]^\sigma \\
M'_\bullet \ar[rr] & & (N'M')_\bullet 
}
\qquad
\xymatrix{
S'_\bullet \ar[r]^\ell \ar[d]_{j'} \ar@/^2em/[rr]^k \ar@/_2em/[dd]_{i'}
         & N'_\bullet \ar[r]^{\psi^{-1}} \ar[d] & N_\bullet \ar[dd] \\
M_\bullet '\ar[r]\ar[d]^{\phi^{-1}} & (N'M')_\bullet \ar@{.>}[dr]^\tau \\
M_\bullet \ar[rr] & & (NM)_\bullet 
}
\]
We now use a standard trick of category theory to see that these unique $\sigma$ and $\tau$ are isomorphisms, and in fact inverses of each other.  Using the uniqueness part of the pushout property we know the identity chain map $(NM)_\bullet {\buildrel 1 \over \longrightarrow} (NM)_\bullet$ is the only morphism for which the diagram
\[
\xymatrix{
S'_\bullet \ar[r]^j \ar[d]_{i'} & N_\bullet \ar[d] \ar@/^/[ddr] \\
M_\bullet \ar[r] \ar@/_/[drr] & (NM)_\bullet \ar@{.>}[dr]^{1} \\
 & & (NM)_\bullet}
\]
commutes.  But the diagram also commutes if the dotted arrow is the map $\tau \circ \sigma$, and therefore $\tau \circ\sigma = 1$.  An identical argument shows that $\sigma\circ\tau$ is the identity chain map $(N'M')_\bullet {\buildrel 1 \over \longrightarrow} (N'M')_\bullet$.  We thus have an isomorphism $\sigma\maps  (NM)_\bullet \to (N'M')_\bullet$ such that the diagram
\[
\xymatrix{
S_\bullet \ar[r] \ar[dr] & (NM)_\bullet \ar[d]_\sigma & S''_\bullet \ar[l] \ar[dl] \\
&                               (N'M')_\bullet
}
\]
commutes.  It is not hard hard to check that $\sigma$ preserves the inner product on $p$-cochains, so that we have $NM\sim N'M'$

Now suppose we have three chain cobordisms 
\[
\xymatrix{
      S_\bullet \ar[r]^i& M_\bullet &  S'_\bullet \ar[l]_{i'}
                }\quad
\xymatrix{
      S_\bullet \ar[r]^j& M'_\bullet &  S''_\bullet \ar[l]_{j'}
                }\quad
\xymatrix{
      S_\bullet \ar[r]^k& M''_\bullet &  S'''_\bullet \ar[l]_{k'}
                }
\]
Showing that composition is associative involves constructing an isomorphism 
\[
          ((M''M')M))_\bullet \buildrel \alpha\over \longrightarrow (M''(M'M))_\bullet 
\]
such that the diagram
\[
\xymatrix{
S_\bullet \ar[r] \ar[dr] & ((M''M')M))_\bullet \ar[d]_\alpha & S'''_\bullet \ar[l] \ar[dl] \\
&                               (M''(M'M))_\bullet
}
\]
commutes, so that the two parenthesizations are equal as equivalence classes.  In fact, $\alpha$ is just the obvious isomorphism.  Checking the details is tedious but not difficult.  In fact, the equivalence relation in the definition of  chain cobordism  was motivated entirely by the need for associativity.

The identity morphism from $S$ to $S$ consists of two identity chain maps
\[
\xymatrix{
      S_\bullet \ar[r]^1& S_\bullet &  S_\bullet \ar[l]_1
                }
\]
which in expanded notation is the diagram
\[
\xymatrix{
0 & S_0 \ar[l] \ar[d]_1 & S_1 \ar[l] \ar[d]_1 & {\cdots} \ar[l] & S_{n-1} \ar[l] \ar[d]_1 \\
0 & S_0 \ar[l] & S_1 \ar[l] & {\cdots} \ar[l] & S_{n-1} \ar[l] & 0 \ar[l] \\
0 & S_0 \ar[u]^{1} \ar[l] & S'_1 \ar[u]^{1} \ar[l] & {\cdots} \ar[l] & S_{n-1} \ar[u]^{1} \ar[l]
}
\]
Here we explicitly write in the position $S_n = 0$ in the middle row just to emphasize that the we are thinking of $S$ both as an {\em object}, an $(n-1)$-complex, and as a {\em morphism}, an $n$-complex.  We write this identity morphism as $1_S\maps  S\to S$.  To see that this does satisfy the left and right  identity axioms, suppose $M\maps S\to S'$, $N\maps S\to S''$.  Then it is a routine exercise to check that the diagrams
\[
\xymatrix{
      S_\bullet \ar[dr] \ar[r] & (MS)_\bullet \ar[d] &  S'_\bullet \ar[dl]\ar[l] \\
      & M_\bullet
                } \qquad \text{and} \qquad 
\xymatrix{
      S'_\bullet \ar[dr] \ar[r] & (SN)_\bullet \ar[d] &  S_\bullet \ar[dl]\ar[l] \\
      & N_\bullet
                }         
\]
commute, where the vertical arrows are the obvious isomorphisms induced by the inclusions of $S$ into $M$ and $N$.

This completes the proof that $\nChain$ is a category.  A {\em monoidal} category is a category with a `tensor product' so we must describe this product in $\nChain$.  In $\nCob$ the tensor product is just the disjoint union of spaces; in $\nChain$ the obvious analog is the direct sum of chain complexes, at both object and morphism levels.  Given objects $S$ and $C$, define $S\oplus C$ to be the chain complex $S_\bullet \oplus C_\bullet$, with inner product on $p$-cochains given by 
\[
 \langle -, - \rangle = \langle -, - \rangle_{S} + \langle -, - \rangle_{C}.
\]
Similarly, for morphisms $M\maps S \to S'$ and $N\maps C \to C'$ define 
\[
M\oplus N\maps S\oplus C \to S'\oplus C'
\]
to be the class of basis preserving and inner product preserving chain maps
\[
S_\bullet \oplus C_\bullet \longrightarrow M_\bullet \oplus N_\bullet 
\longleftarrow S'_\bullet \oplus C'_\bullet.
\]
The identity object for $\oplus$ is the trivial chain complex $0_\bullet$:
\[
  \xymatrix{ 0 & 0 \ar[l] & {\cdots} \ar[l] & 0 \ar[l] }
\]
and the associator and unit laws are inherited from the obvious abelian group isomorphisms $(G\oplus H)\oplus K \iso G\oplus (H\oplus K)$ and $G\oplus 0\iso G \iso 0\oplus G$.  

Making $\nChain$ a {\em symmetric} monoidal category requires that we also specify a `symmetry' or `braiding' which lets us switch the order in the tensor product.  This comes from the abelian group isomorphisms $G\oplus H \iso H \oplus G$ in an obvious way.  Checking that these constructions actually yield a symmetric monoidal category involves checking certain `coherence laws' guaranteeing the associator, unit laws and braiding get along appropriately.  This a lengthy process, but not a difficult one.  We refer the reader to \cite{MacLane} for the detailed definitions of monoidal and symmetric monoidal categories. $\blacksquare$

\begin{thm}
\label{chainFT}
Lattice $p$-form electromagnetism gives a chain field theory.
\end{thm}

Before proving this theorem, we remark that for cobordisms we must modify our definition of the {\bf action} slightly.  The idea of the modified action we shall use for cobordisms $M\maps S\to S'$ is that the action on $p$-cells in the `boundary' $S+S'$ should only count {\em half} as much as the action on cells in the interior.  This keeps us from `overcounting' the action when we compose two cobordisms $S\buildrel M \over \longrightarrow S' \buildrel M' \over \longrightarrow S''$.   If $M\maps S\to S'$ is a chain cobordism, let us rename our old na\"ive action $\tilde S$, so that
\[
   \tilde S(A) = -\frac{1}{2e^2} h^{ij}F(x_i) F(x_j)
\]
in the $\R$ case.  If $A$ is a $p$-connection on the spacetime cobordism $M\maps S \to S'$, we then define the action on $M$ to be
\[
 S(A) = \tilde S(A) - \half \tilde S(A|_S) - \half \tilde S(A|_{S'}).
\]

In the $U(1)$ case, we don't define the action, but only its `exponential' --- the theta function in equation (\ref{u1S}) in Section \ref{u1path}.
\begin{equation*}
  e^{-\tilde S(A)} = 
     \sum_{n\in \Z^{N}} e^{-\frac{1}{2e^2} h^{ij} (A(\partial x_i) - 2n_i \pi)(A(\partial x_j) - 2n_j \pi)}
\end{equation*}
This leads us to define for a $p$-connection $A$ on $M\maps S \to S'$,  
\[
  e^{-S(A)} := e^{-\tilde S(A)} e^{\half \tilde S(A|_S)} e^{\half \tilde S(A|_{S'})}.
\]   
where, $e^{\half \tilde S}$ just denotes the obvious thing---the reciprocal of the square root of the theta function ``$e^{-S}$". 

\vspace{.8em}

\noindent{\bf Proof of Theorem \ref{chainFT}:}  To define a functor
\[
         Z\maps  \nChain \to \Hilb
\]
we must specify what $Z$ does to objects and morphisms.  When space is the $(n-1)$-complex $S$, the classical configuration space of $p$-form electromagnetism with gauge group $U(1)$ is the space of $p$-connections on $S$ modulo gauge transformations:
\[
   \frac{{\cal A}(S)}{{\cal G}(S)} = \frac{C^p(S,U(1))}{B^p(S,U(1))}
\]
Quantization gives the Hilbert space of states, the space of all square-integrable functions on classical field configurations.  We thus define
\[
          Z(S) := L^2\left(\frac{{\cal A}(S)}{{\cal G}(S)}\right).
\]

Now $Z$ must also assign to each morphism $M\maps S \to S'$ in $\nChain$ a linear map $Z(M)\maps Z(S) \to Z(S')$ between Hilbert spaces, corresponding to `time evolution' of states on the  slice $S$ of space to the slice $S'$.  To determine $Z(M)$ it suffices to  specify, for any $\psi\in Z(S) =L^2({\cal A}(S)/{\cal G}(S))$  and $\phi\in Z(S') =L^2({\cal A}(S')/{\cal G}(S'))$ the transition probability
\[
 \langle \phi, Z(M) \psi \rangle =
      \int_{{\cal A}(M)} \bar \phi(A |_{S'})\psi(A |_{S}) e^{-S(A)} {\cal D}A
\]
where for convenience we absorb the normalization factor into the definition of the measure ${\cal D}A$ --- we can do this since the partition function converges when the gauge group is $U(1)$.   Also because we are using $U(1)$, factoring out gauge transformations is not necessary for the path integral to make sense. Hence we integrate over ${\cal A}(M)$ rather than ${\cal A}(M)/{{\cal G}(M)}$.  Since $U(1)$ is compact and the integrand is gauge invariant, the result is the same. 

To show functoriality we must show that this definition of $Z$ respects identity morphisms and composition.  First suppose we apply $Z$ to the identity cobordism $1_S\maps S\to S$, i.e.\
\[
\xymatrix{
       S_\bullet \ar[r]^{1} & S_\bullet & S_\bullet \ar[l]_{1} 
                 }
\] 
Then given two states $\psi, \phi \in Z(S) = L^2({\cal A}(S)/{\cal G}(S))$ we have 
\begin{align*}
 \langle \phi, Z(1_S) \psi \rangle &=
      \int_{{\cal A}(S)}\bar \phi(A|_S)\psi(A |_S) e^{-S(A)} {\cal D}A \\
     &= \int_{{\cal A}(S)}\bar \phi(A|_S)\psi(A |_S)  {\cal D}A
                               = \langle \phi,\psi \rangle 
\end{align*}
where we have used the fact that for the identity cobordism $1_S$, we have
\[
   e^{-S(A)} = e^{-\tilde S(A)} e^{\half \tilde S(A)} e^{\half \tilde S(A)} = 1
\]
Hence $Z(S)$ is the identity on $Z(S) = L^2({\cal A}(S)/{\cal G}(S))$.  

As usual, showing that $Z$ respects composition is the nontrivial part of showing $Z$ is a functor.  Suppose we have two composable morphisms in $\nChain$:
\begin{align*}
                    M\maps & S \to S' \\
                    M'\maps & S' \to S''.
\end{align*}
Since $M'M$ is a pushout, the diagram
\[
\xymatrix{
S'_p \ar[r]^{} \ar[d]_{} & M_p \ar[d] \ar@/^/[ddr]^{A} \\
M'_p \ar[r] \ar@/_/[drr]^{A'} & (M'M)_p \ar@{.>}[dr]^{A^\circ} \\
 & & U(1)}
\]
shows that a $p$-connection $A^\circ$ on the composite $M'M$ is precisely a $p$-connection $A$ on $M$ together with a $p$-connection $A'$ on $M'$ such that $A'$ and $A$ agree when restricted to the common boundary $S'$.  To enforce this agreement on $S'$ in path integrals, we will use a delta function $\delta( A|_{S'}- A'|_{S'})$.  If $  \{ \eta_\lambda : \lambda \in \Lambda \} $
is an orthonormal basis of $L^2({\cal A}(S')/{\cal G}(S'))$, then expanding our delta function gives
\begin{equation}
\label{delta}
\begin{split}
\delta( A|_{S'}- A|_{S'}) 
&=\sum_{\lambda\in \Lambda}
        \left\langle \eta_\lambda(A'|_{S'}), \delta( A|_{S'}- A'|_{S'})\right \rangle \,\eta_\lambda(A' |_{S'}) \\
              &=
      \sum_{\lambda\in \Lambda}
      \left( \int  \bar \eta_\lambda(A'|_{S'}) \delta( A|_{S'}- A'|_{S'})  {\cal D}A'\right) 
           \eta_\lambda(A' |_{S'})   \\
           &=  
      \sum_{\lambda\in \Lambda}
       \bar \eta_\lambda(A|_{S'})  \eta_\lambda(A' |_{S'})          
\end{split}
\end{equation}

Now to show $Z(M'M) = Z(M')Z(M)$, it suffices to show
\[
     \langle \phi, Z(M'M)\psi \rangle = \langle \phi, Z(M')Z(M)\psi \rangle
\]
for all $\psi \in Z(S)$ and $\phi \in Z(S'')$.  We have:
\begin{align*}
  \langle \phi, Z&(M')Z(M)\psi \rangle \\ 
  \qquad &= 
      \int_{{\cal A}(M')} \bar \phi(A' |_{S''})(Z(M)\psi)(A' |_{S'}) e^{-S(A')} {\cal D}A' \\
              &=
      \int_{{\cal A}(M')} \bar \phi(A' |_{S''})
               \sum_{\lambda\in \Lambda}\langle \eta_\lambda, Z(M)\psi\rangle\eta_\lambda(A' |_{S'}) e^{-S(A')} {\cal D}A' \\
              &=
      \int_{{\cal A}(M')} \int_{{\cal A}(M)} \bar 
      \phi(A' |_{S''})\psi(A|_{S})
               \sum_{\lambda\in \Lambda} \bar \eta_\lambda(A|_{S'}) \eta_\lambda(A' |_{S'}) e^{-S(A)} e^{-S(A')} 
               {\cal D}A{\cal D}A' \\  
              &=
      \int_{{\cal A}(M'M)} \bar 
      \phi(A^\circ|_{S''})\psi(A^\circ|_{S}) e^{-S(A^\circ)}        {\cal D}A^\circ \\
      &= \langle \phi, Z(M'M)\psi \rangle,
\end{align*}
as desired.  
Here we have used the expansion (\ref{delta}) and the fact that
\begin{align*}
  e^{-S(A)}e^{-S(A')} = e^{-S(A^\circ)}.
\end{align*}   
This is comes from the definition of the action on a cobordism in the case where the gauge group is $\R$.  Using this and the definition (\ref{comp-ip}) of the inner product for a composite of cobordisms, we obtain: 
\begin{align*}
  S(A) + S(A') &= \tilde S(A) + \tilde S(A') 
                             - \half \left(\tilde S(A|_{S'}) + \tilde S(A'|_{S'})\right) 
                             - \half \tilde S(A|_S) - \half \tilde S(A'|_{S''})  \\
                         &= \tilde S(A^\circ|_M) + \tilde S(A^\circ|_{M'}) 
                             - \tilde S(A^\circ|_{S'})  
                             - \half \tilde S(A^\circ|_S) - \half \tilde S(A^\circ|_{S''})  \\
                         &= \tilde S(A^\circ) 
                             - \half \tilde S(A^\circ|_S) - \half \tilde S(A^\circ|_{S''})  \\
                         &  = S(A^\circ).
\end{align*}   
The result follows in the case where the gauge group is $U(1)$ from the relationship between Gaussians and theta functions.  This completes the proof that $Z$ is a functor.

The essential features of a {\em symmetric monoidal} functor are a natural isomorphism
\[
         \Phi\maps Z(-)\tensor Z(-) \stackrel \sim \Longrightarrow Z(-\oplus -)
\]
between the functors
\begin{align*}
            Z(-)\tensor Z(-)&\maps  \nChain \times \nChain \to \Hilb \\
       \text{and }\;      Z(-\oplus -)&\maps  \nChain \times \nChain \to \Hilb
\end{align*}
and an isomorphism
\[
               \phi \maps  1_\Hilb \stackrel \sim \longrightarrow Z(0_{\nChain}) 
\]
between the identity in $\Hilb$ and $Z$ applied to the identity in $\nChain$.  The existence of $\phi$ is obvious: $Z(0)$ is the space of $L^2$ functions on a one-point set, i.e.\ complex numbers, and $\C$ is the identity for the tensor product of Hilbert spaces.  To complete the proof that the functor is symmetric monoidal, one must work out what $\Phi$ is and check that $\Phi$ and $\phi$ satisfy the required coherence laws.  We leave these details to the reader, whom we again refer to Mac Lane's textbook \cite{MacLane} for the definitions. $\blacksquare$

\section*{Appendix: Gaussian Integration}
\addtocontents{toc}{
\contentsline {section}{\numberline {}Appendix: Gaussian Integration}{\thepage}}
The fundamental integral is the definite integral of the Gaussian function $e^{-x^2/2}$ over all of $\R$:
\[
\int_{-\infty}^{\infty} e^{-x^2/2} dx =\sqrt{2\pi}.
\]
More generally, if we consider the Gaussian $e^{-x^2/2\sigma^2}$, with {\bf standard deviation} $\sigma$, a simple change of variables shows that
\[
\int_{-\infty}^{\infty} e^{-x^2/2\sigma^2} dx =\sqrt{2\pi} \sigma.
\]
This gives us the usual normalizing constant for a one-dimensional Gaussian probability distribution.  Thus, using the properly normalized measure,
\[
   \frac{e^{-x^2/2\sigma^2}\,dx}{\sqrt{2\pi}\sigma},
\]
we can calculate expected values of random variables. Given a function $f$ of the Gaussian random variable $x$, the expected value of $f$ is given by:
\[
\langle f \rangle = \frac{1}{\sqrt{2\pi} \sigma}
 \int_\R f(x)e^{-x^2/2\sigma^2}\,dx
\] 
In general, such integrals are of course hard to carry out.  Since any reasonably nice function can be approximated by polynomials, however, we will be content to calculate $\langle x^n\rangle$.  It is not hard to show, using integration by parts and the fact that odd functions integrate to zero, that 
\[
\langle x^n \rangle := 
 \frac{1}{\sqrt{2\pi} \sigma}
 \int_{-\infty}^{\infty} x^{n} e^{-x^2/2\sigma^2} dx
 =\left\{
  \begin{array}{cl}
     \sigma^{n}(n-1)!! & \textrm{$n$ even}\\
      0                             & \textrm{$n$ odd}
  \end{array}\right.
.
\]
Here, the double factorial $(n-1)!!$ means $(n-1)\cdot(n-3)\cdots (5)\cdot(3)\cdot(1)$, and we evidently need to define $(-1)!!=1$ for the formula to be consistent with our original Gaussian integral.

Next we generalize Gaussian integration to more than one dimension.  The simplest case is the totally symmetric Gaussian bump, which has an obvious solution:
\[
 \int_{-\infty}^\infty
  \cdots\int_{-\infty}^\infty
      e^{-\frac{1}{2}(x_1^2+ \cdots +x_n^2)}
  dx_1 \cdots dx_n 
   = \prod_{k=1}^{n} \sqrt{2\pi}
   = (2\pi)^{n/2},
\]
which we can also write using the standard inner product on $\R^n$ as
\[
 \int_{\R^n} e^{-\frac{1}{2}\langle x,x \rangle} d^n x 
       = (2\pi)^{n/2}.
\]
Often, however, we want to do similar integrals where the argument of the exponential is some more general quadratic function of the $x_i$.  
It is a basic result of linear algebra that any quadratic form $q\maps V \to \R$ on an inner product space $V$ may be written as
\[
q( x)= \langle  x , Q x \rangle
\]
where $Q$ is a self-adjoint linear operator $V\rightarrow V$, i.e.\ a symmetric matrix\footnote{In a more general setting, a self-adjoint linear operator is represented by a {\em hermitian} matrix, but since in this paper we are concerned with real vector spaces, the adjoint is merely the transpose.}.  The matrix $Q$ is called positive whenever the quadratic form $q$ is positive.

We now calculate
\[
   \int_{\Rn} e^{-\frac{1}{2}\langle x,Qx \rangle} d^{n}x
\]
where $Q$ is a symmetric, positive definite $n\times n$ matrix. To do this, note that the positivity of $Q$ implies $Q$ has a unique positive square root --- this is just the finite dimensional version of the square root lemma of functional analysis (cf.\ Reed and Simon \cite{RS} p.\ 196).  That is, there is a positive definite matrix $B$ with \(B^2=Q\).  Moreover, since $Q$ is symmetric, so is B, so we can write 
\[
  \langle x,Qx \rangle =
  \langle x,BBx \rangle =
  \langle B^\dagger x,Bx \rangle =
  \langle Bx,Bx\rangle,
\] 
where the adjoint $B^\dagger$ in the present case is simply the matrix transpose.  To do the above integral, we can thus make the change of variables $y=Bx$, or in component form $y^i = B^i_j x^j$. Using the proper Jacobian for the transformation, and the fact that $\det(B)=\sqrt{\det(Q)}$, we find
\begin{align*}
   \int_{\Rn} e^{-\frac{1}{2}\langle x,Qx \rangle} d^{n}x
   &=\sqrt{\frac{(2\pi)^n}{\det(Q)}}.
\end{align*}
To calculate expected values: If 
\[
    \ell_1,\ell_2,\ldots, \ell_m \maps  \R^n \to \R
\]
are $m$ (not necessarily distinct) linear functionals on $\R^n$ then we have the following simple formula for calculating the expected value of their product:
\begin{align*}
\langle \ell_1 \ell_2 \ldots \ell_m \rangle &=
\sqrt{\frac{\det(Q)}{(2\pi)^n}}
\int_{\Rn} \ell_1(x) \ell_2(x) \ldots \ell_m(x) e^{-\frac{1}{2}\langle x,Qx \rangle} d^{n}x \\
&=
\sum_\text{pairings}\prod_{\text{ pairs }\ell_i, \ell_j} \langle \ell_i  Q^{-1}, \ell_j \rangle
\end{align*}
The proof of this relies on the facts that both sides of this expression are linear in each of the $\ell_i$, and two multilinear functionals agree if and only if they agree when all of the linear factors are the same.

In particular, for $m=2$, if we take $Q$ to be the identity on $\R^n$ so that we are back to the symmetric Gaussian bump, we have a formula which, aside from the comma on the right side, looks like a tautology: 
\[
      \langle \ell_1 \ell_2 \rangle = \langle \ell_1,\ell_2 \rangle.
\]
For $m=4$ we have
\[
     \langle \ell_1 \ell_2 \ell_3 \ell_4 \rangle = 
           \langle \ell_1,\ell_2 \rangle \langle \ell_3,\ell_4 \rangle +
           \langle \ell_1,\ell_3 \rangle \langle \ell_2,\ell_4 \rangle +
           \langle \ell_1,\ell_4 \rangle \langle \ell_2,\ell_3 \rangle
\]
where the $3$ terms correspond to the $3= (4-1)!!$ distinct pairings of $\ell_1, \ldots,\ell_4$:
\[
\xy
(-30,0)*{\xy
	     (-5,5)*{\ell_1}="A";
	     (5,5)*{\ell_2}="B";
	     (-5,-5)*{\ell_3}="C";
	     (5,-5)*{\ell_4}="D";
	     "A"+(2,0);"B"+(-2,0)**\dir{-};
	     "C"+(2,0);"D"+(-2,0)**\dir{-};
         \endxy};
(0,0)*{\xy
	     (-5,5)*{\ell_1}="A";
	     (5,5)*{\ell_2}="B";
	     (-5,-5)*{\ell_3}="C";
	     (5,-5)*{\ell_4}="D";
	     "A"+(0,-2);"C"+(0,2)**\dir{-};
	     "B"+(0,-2);"D"+(0,2)**\dir{-};
         \endxy};
(30,0)*{\xy
	     (-5,5)*{\ell_1}="A";
	     (5,5)*{\ell_2}="B";
	     (-5,-5)*{\ell_3}="C";
	     (5,-5)*{\ell_4}="D";
	     "A";(-.5,.5)**\dir{-};
	     (.5,-.5);"D" **\dir{-};
	     "B";"C"**\dir{-};
         \endxy};
\endxy
\]
Note that the number of distinct pairings of $m$ objects is zero when $m$ is odd and $(m-1)!!$ when $m$ is even, so in the one-dimensional case where each of the linear functionals is just the identity map $x\mapsto x$ and $Q$ is just the number $1/\sigma^2$, the formula reduces to 
\[
   \langle x^m \rangle = 
   \left\{
  \begin{array}{cl}
     \sigma^{n}(n-1)!! & \textrm{$n$ even}\\
      0                             & \textrm{$n$ odd}
  \end{array}\right .
\]
as we found before.

\section*{Acknowledgements}  
I thank John Baez for numerous discussions on this subject.  This paper would not have been written without all I learned from John before and during its writing.  I also thank Jeffrey Morton, Hendryk Pfeiffer, and Urs Schreiber for useful discussions or correspondence.  Some of the early conversation which led to this work took place on the newsgroup sci.physics.research, and I wish to acknowledge those who followed along and participated.  Among them, I must single out Eric Forgy as having been particularly enthusiastic and helpful.

\end{document}